\documentclass[prd,twocolumn,balance,floatfix,superscriptaddress,nofootinbib]{revtex4-1}
\usepackage{dcolumn,amsmath}
\usepackage{graphicx}
\usepackage{braket}
\usepackage{bm}
\usepackage{amssymb}
\usepackage{hyperref}
\usepackage{multirow}
\usepackage{footnote}
\usepackage{subcaption}
\usepackage{ragged2e}
\usepackage{xcolor}
\usepackage{pdfcomment}
\usepackage[justification=justified,labelfont=bf,singlelinecheck=off]{caption}
\usepackage{threeparttable,booktabs}  
\usepackage{subcaption,siunitx,booktabs}
\usepackage{caption,afterpage,tabularx}
\usepackage{diagbox}
\usepackage{mathrsfs}
\usepackage{cancel}
\usepackage{amsmath}
\usepackage{caption}
\usepackage{subcaption}

\usepackage{float}
\usepackage{mwe}
\usepackage[title]{appendix}
\usepackage{balance}

\newcommand{\JSU}{
School of Physics and Electronic Engineering,
\\Jiangsu University, Zhenjiang, 212013 Jiangsu, China\\
}

\newcommand{\MUST}{State Key Laboratory of Lunar and Planetary Sciences, 
\\Macau University of Science and Technology, 999078 Macao, China}

\begin{document}

\title{Neutron-antineutron oscillation accompanied by CP-violation in magnetic fields}

\author{Yongliang Hao}
\email{yhao@ujs.edu.cn}
\affiliation{\JSU}
\affiliation{\MUST}
\author{Kamphamba Sokalao Nyirenda}
\affiliation{\JSU}
\author{Zhenwei Chen}
\affiliation{\JSU}


\date{\today}

\begin{abstract}

In this work, we explore the possibility of the $n$-$\bar{n}$ oscillation accompanied by CP-violation in the presence of magnetic fields. The $n$-$\bar{n}$ oscillation, which violates the baryon number ($\mathcal{B}$) by two units ($|\Delta \mathcal{B}| = 2$), can originate from the mixing between the neutron ($n$) and the neutral elementary particle ($\eta$) and may give rise to non-trivial physical consequences that can be testable in future experiments. We show that the probability of the $n$-$\bar{n}$ oscillation can be greatly enhanced by properly adjusting the magnetic field. In particular, the peak values of the oscillation probability in the presence of resonance magnetic fields can be $8$-$10$ orders of magnitude higher than that in the absence of magnetic fields. We point out that there might not be sizable CP-violating effects in the $n$-$\bar{n}$ oscillation unless the mass of $\eta$ is close to the mass of the neutron. We also analyze the interplay between various parameters associated with both $\mathcal{B}$-violation and CP-violation and attempt to disentangle the effects of such parameters. The $n$-$\bar{n}$ oscillation process accompanied by CP-violation may open a promising avenue for exploring new physics beyond the Standard Model (SM).

\end{abstract}

\maketitle

\section{Introduction\label{sec1}}

Cold and ultra-cold neutron sources can serve as rich and varied environment where many interesting processes can occur \cite{snow2022searches,alarcon2023fundamental}, allowing for the search for new physical phenomena beyond the SM in smaller experiments compared with the ones at the LHC. In contrast to charged particles, the experiments with neutrons face more difficulties in both the acceleration and manipulation of neutrons \cite{dubbers2011neutron}, leading to more uncontrollable uncertainties. Due to such limitations, there are still many puzzles about neutrons that have not been well-explained, such as the neutron lifetime anomaly (see e.g. Refs. \cite{yue2013improved, gonzalez2021improved,particle2022review,tan2023neutron}) and various interpretations of the experimental results of the $n$-$\bar{n}$ oscillation (see e.g. Refs. \cite{mohapatra2009neutron,phillips2016neutron}).

The $n$-$\bar{n}$ oscillation, which violates the baryon number ($\mathcal{B}$) by two units ($|\Delta \mathcal{B}|=2$), has received considerable attention both theoretically and experimentally \cite{mohapatra2009neutron,phillips2016neutron}. The experimental searches for the $n$-$\bar{n}$ oscillation have been performed in various mediums \cite{baldo1994new,jones1984search,takita1986search, berger1990search,chung2002search,aharmim2017search,abe2015search,abe2021neutron}, including bound states, field-free vacuum, etc. Up to date, no significant signal for the $n$-$\bar{n}$ oscillation has been found \cite{baldo1994new,jones1984search,takita1986search, berger1990search,chung2002search,aharmim2017search,abe2015search,abe2021neutron}. In field-free vacuum, the lower limit on the $n$-$\bar{n}$ oscillation time reported by the Institut Laue-Langevin (ILL) experiment is approximately $0.86 \times 10^{8}$ s \cite{baldo1994new}. In bound states,  $n$-$\bar{n}$ oscillation is greatly suppressed by environmental factors due to the difference in the potential which neutrons and antineutrons feel (see e.g. Refs. \cite{sandars1980neutron,alberico1982new,dover1983neutron}). The searches for the $n$-$\bar{n}$ oscillation in bound states have been performed by various experiments, such as Irvine-Michigan-Brookhaven (IMB) \cite{jones1984search}, Kamiokande (KM) \cite{takita1986search}, Frejus \cite{berger1990search}, Soudan-2 (SD-2) \cite{chung2002search}, Sudbury Neutrino Observatory (SNO) \cite{aharmim2017search}, Super-Kamiokande (Super-K) \cite{abe2015search,abe2021neutron}, etc. Among them, the most stringent constraint on the $n$-$\bar{n}$ oscillation time is imposed by the Super-K experiment with a value of $4.7 \times 10^{8}$ s \cite{abe2021neutron}, when converted to the field-free vacuum value.

The possible interaction between neutron ($n$) and mirror-neutron ($n^{\prime}$) may give rise to new-physics effects (see e.g. Ref. \cite{addazi2021new}), such as neutron disappearance ($n$-$n^{\prime}$) \cite{berezhiani2006neutron,berezhiani2009more,berezhiani2012magnetic,berezhiani2019neutron2,addazi2021new,berezhiani2021neutron,berezhiani2019on,babu2022theoretical,tan2023neutron} and neutron regeneration  [$n$-$n^{\prime}$($\bar{n}^\prime$)-$n$] \cite{berezhiani2006neutron,pokotilovski2006experimental,berezhiani2009more,berezhiani2017neutron,addazi2021new,berezhiani2019on,kamyshkov2022neutron}. Furthermore, the $n$-$\bar{n}$ oscillation may also occur indirectly through the interaction between $n$ and $n^{\prime}$, i.e. [$n$-$n^{\prime}$($\bar{n}^\prime$)-$\bar{n}$] \cite{mohapatra2005some,berezhiani2016neutron,addazi2021new,berezhiani2021possible,kamyshkov2022neutron}. The experimental searches for the $n$-$\bar{n}$ oscillation mediated by $n^{\prime}$ have been designed \cite{addazi2021new,ayres2022improved} and preliminary results of the relevant $n$-$n^{\prime}$ oscillation have been reported \cite{broussard2022experimental}. The measurement can be accomplished by a multi-stage program \cite{addazi2021new,acharya2023fundamental}. In the measurement, a homogeneous magnetic field can be applied and its magnitude can be changed in a small increment for each step so that a resonance effect, if it exists, can be captured \cite{berezhiani2021possible}. As a relevant process, the preliminary results of the $n$-$n^{\prime}$ oscillation have been reported, suggesting that the $n$-$n^{\prime}$ oscillation as an explanation for the neutron lifetime anomaly in magnetic fields as large as $6.6$ T can be excluded for a specific range of mass difference ($\delta m \equiv m_{n^{\prime}} - m_n$) \cite{broussard2022experimental}. However, the whole process of the $n$-$\bar{n}$ oscillation mediated by $n^{\prime}$ remains experimentally unexplored \cite{perez2022baryon}.  
Motivated by such studies \cite{mohapatra2005some,berezhiani2016neutron,addazi2021new,berezhiani2021possible,kamyshkov2022neutron}, the $n$-$\bar{n}$ oscillation induced by the mixing between the neutron and the elementary neutral particle ($\eta$) has also been suggested \cite{hao2022neutron}.

CP-violation makes it possible to absolutely distinguish between matter and antimatter \cite{branco1999cp,sozzi2007discrete,bigi2009cp}. CP-violation was first observed in the neutral $K$ meson systems \cite{christenson1964evidence}. In the neutral $B$ meson systems, the observability of CP-violation was suggested by Bigi and Sanada \cite{bigi1981notes}, and was observed by the BABAR \cite{aubert2001observation} and Belle \cite{abe2001observation} experiments. The evidence of CP-violation in the $B_s^0$ meson systems has also been reported by the LHCb experiment \cite{aaij2013first}. In the $D^{0}$ meson systems, CP-violation has been reported by the LHCb experiment with an asymmetry of $-15.4 \pm 2.9$ $(\text{stat. + syst.}) \times 10^{-4}$, which is different from zero at the level of $5.3$ $\sigma$ \cite{aaij2019observation}. Recently, a possible signal of CP violation in the neutrino systems was also reported by the T2K experiment \cite{t2k2020constraint,*t2k2020publisher,abe2023measurements}.

$\mathcal{B}$-violation and CP-violation (together with the C-violation) are two of
the important conditions suggested by Sakharov to explain the observed matter-antimatter asymmetry in our universe \cite{sakharov1967violation,*sakharov1967violation2}. Furthermore, since $\mathcal{B}$-violation and CP-violation are usually implemented as important features in many new physics models \cite{particle2022review}, they play a critical role in testing the SM and in constructing new physics models beyond the SM. It is predicted that CP-violation can possibly occur in the $n$-$\bar{n}$ oscillation process mediated by an elementary neutral particle $\eta$\footnote{The symbol $\eta$ represents a new elementary particle outside the SM and should not be confused with the isosinglet meson.} \cite{hao2022neutron}. The $n$-$\bar{n}$ oscillation accompanied by observable CP-violating effects may open a promising avenue for explaining the origin of the matter-antimatter asymmetry and could provide an appealing scenario for exploring new physics beyond the SM.

This paper is organized as follows. To begin with, we briefly review the model that leads to the $n$-$\bar{n}$ oscillation. Next, we review the mathematical description of the $n$-$\bar{n}$ oscillation mediated by $\eta$ and its connection with CP-violation. Then, we transfer our attention to the electromagnetic (EM) properties of the neutron and in particular to the impact of magnetic fields on the $n$-$\bar{n}$ oscillation probability. After that, we explore the possible enhancement effect of the $n$-$\bar{n}$ oscillation by adjusting the magnitude of the magnetic field. In addition to this, we also analyze the interplay among the various parameters associated with both $\mathcal{B}$-violation and CP-violation, and attempt to disentangle their effects. Finally, we choose the experimental results of CP-violation in the meson systems as representative benchmark values of CP-violation and discuss the observability of CP-violation in the $n$-$\bar{n}$ oscillation process. We also show that such an analysis would provide helpful clues and insights into future studies of new physics beyond the SM. In the following discussions, unless otherwise specified, we will adopt the natural units (i.e. $c \equiv 1$, $\hbar \equiv 1$).

\section{The model \label{nnmixing}}

Although neutrons, in a brief picture, contain charged particles (e.g. quarks), they seem electrically neutral as a whole with very high precision to the outside \cite{particle2022review}. In reality, neutrons can interact with the EM fields through the anomalous magnetic moment \cite{bjorken1964relativistic,mckeen2016c}. The mixing between $n$ and $\eta$ violates both the baryon number ($\mathcal{B}$) and the lepton number ($\mathcal{L}$) by one unit ($|\Delta \mathcal{B}| = 1$, $|\Delta \mathcal{L}| = 1$) but conserves their difference $[|\Delta(\mathcal{B}-\mathcal{L})|=0$]. Here, $\eta$ may have a non-zero lepton number ($\mathcal{L} \equiv 1$) and have similar properties with sterile neutrinos. If $\eta$ is stable, it might be a dark matter candidate, otherwise its decay products might be dark matter candidates. Such a mixing may give rise to non-trivial physical consequences that are testable in future experiments. The $n$-$\bar{n}$ oscillation\footnote{The $n$-$\bar{n}$ oscillation can also be predicted by other new-physics models, such as the extra dimensional model (see e.g. Refs. \cite{nussinov2002n,girmohanta2020nucleon}).} is one of the interesting physical consequences resulting from the $n$-$\eta$ mixing and can be described by the following effective Lagrangian (density) \cite{fornal2018dark,fornal2023neutron}:
\begin{equation}
\begin{split}
\mathscr{L}_{\text{eff}}  \supset & -\frac{1}{4} F_{\mu\nu}F^{\mu\nu} + \bar{\eta} (i \cancel{\partial} -m_{\eta}) \eta + \bar{n} \Big(i \cancel{\partial} -m_{n}\Big) n \\ 
& + \delta(\bar{n}\eta + \text{H.c.}) + \frac{g_n}{2m_n}\bar{n} \sigma_{\mu \nu} n F^{\mu \nu}.
\end{split}
\end{equation}
The first term represents the kinetic energy of the EM fields, where $F^{\mu\nu}$ is the EM field strength tensor. The second term represents the interaction-free energy of $\eta$ including a kinetic energy term and a mass term. The third term represents the interaction-free energy of $n$ including a kinetic energy term and a mass term. The fourth term represents the $n$-$\eta$ mixing term \cite{mckeen2018neutron}. The fifth term represents the interaction between the neutron and the EM field \cite{mckeen2016c}. The g-factor of the neutron takes the value: $g_n \simeq -3.826$ \cite{tiesinga2021codata,tiesinga2021codata2}. $m_n$ and $m_{\eta}$ are the masses of $n$ and $\eta$, respectively. Since $\eta$ is a new particle outside the SM and is closely related to dark matter, it is quite natural to assume that the magnetic dipole moment of $\eta$ is negligible so that it barely interacts with EM fields and thus there is no direct coupling between $\eta$ and the EM field.

Relevant studies have provided clues to place constraints on $m_{\eta}$. To kinematically forbid proton decay and $\eta$ decay, the following constraint can be imposed on $m_{\eta}$: $m_p -m_e \lesssim m_{\eta} \lesssim m_p +m_e$ (see e.g. Refs. \cite{allahverdi2013natural,allahverdi2014kev,allahverdi2018simple,jin2018nucleon,fajfer2021colored,mckeen2023long}), where $m_{p}$ and $m_{e}$ are the masses of the proton and electron, respectively. In order to stabilize the $^9$Be nuclei, the constraint $m_{\eta}> 937.9$ MeV can be further imposed \cite{mckeen2016c,mckeen2018neutron}. Such restrictions were obtained indirectly from experiments and more or less depend on specific theoretical models. Although the restrictions are very stringent, there is a lack of direct experimental support. Alternatively, $m_{\eta}$ might be located outside this narrow range while the stability of proton and nuclei can be guaranteed by imposing an additional assumptions and symmetries, e.g. $Z_2$-symmetry \cite{gu2011baryogenesis,dev2015tev}. Furthermore, another possibility that has not been excluded is that $\eta$ might be unstable and thus not be a dark matter candidate, but instead its decay products might be stable and thus could be dark matter candidates. Therefore, we think it might be interesting to relax the conditions on $m_{\eta}$ to the whole range below $m_n$, making it possible to explore the impact of small $m_{\eta}$ on the physical properties of interest (e.g. CP-violation) more thoroughly. Therefore, we assume that $m_{\eta}$ satisfies the condition: $m_{\eta} \lesssim m_n$.

Since the choice of experimental data and theoretical models is different, the phenomenological constraints on the mass of exotic particles vary remarkably and, to some extent, contradict each other, making it difficult to compare them. For example, previous study shows that the existence of neutron stars (NS) with mass values greater than $0.7$ $M_{\odot}$ tends to require the exotic baryonic particles either to be heavier than $1.2$ GeV or to have strong repulsive self-interactions \cite{mckeen2018neutron}. We point out that our assumptions on $m_{\eta}$ are consistent with such constraints, because the neutral particle $\eta$ under discussion in this work is a leptonic particle, rather than a baryonic particle. We assume that $\eta$ has similar properties with sterile neutrinos and barely interacts with the ordinary matter. In this case, $\eta$ can escape from the interior of a neutron star nearly without any collisions, making it unlikely to accumulate inside the neutron star and thus unlikely to have a significant impact on the chemical composition and equation of state (EoS) of the neutron star. Therefore, the constraints imposed by neutron stars may not be applicable to $\eta$.

In the presence of external magnetic fields, the probabilities of the $n$-$\bar{n}$ and $\bar{n}$-$n$ oscillations can be given by\footnote{Note the direction of the magnetic field in Eq. (\ref{fullpnn2}) is opposite to that in Eq. (\ref{fullpnn1}). In Eq. (\ref{fullpnn1}), $\theta_1$ ($\theta_2$) is the mixing angle associated with the $n$-$\eta$ ($\bar{n}$-$\bar{\eta}$) mixing, while in Eq. (\ref{fullpnn2}), $\theta_1$ ($\theta_2$) is the mixing angle associated with the $\bar{n}$-$\bar{\eta}$ ($n$-$\eta$) mixing.} \cite{hao2022neutron}
\begin{equation}
\begin{split}
P_{n \rightarrow \bar{n}}^{B \neq 0} \simeq & \sin^2(2\theta_1)\sin^2\Big(\frac{\phi_1}{2}\Big) \Big[\Big(\frac{m_\eta}{m_n}\Big)^2 \cos^4\theta_1 + \sin^4\theta_1 \\
&+ \frac{1}{2} \Big( \frac{m_\eta}{m_n} \Big)  \sin^2(2\theta_1) \cos(\phi_1 + 2\xi )\Big] \sin^2(2\theta_2)\\
&\times\sin^2\Big(\frac{\phi_2}{2}\Big).    
\end{split}
\label{fullpnn1}
\end{equation}
\begin{equation}
\begin{split}
P_{\bar{n} \rightarrow n}^{B \neq 0} \simeq & \sin^2(2\theta_1)\sin^2\Big(\frac{\phi_1}{2}\Big) \Big[\Big(\frac{m_\eta}{m_n}\Big)^2 \cos^4\theta_1 + \sin^4\theta_1\\
&+ \frac{1}{2} \Big( \frac{m_\eta}{m_n} \Big)  \sin^2(2\theta_1) \cos(\phi_1 - 2\xi )\Big] \sin^2(2\theta_2)\\
&\times\sin^2\Big(\frac{\phi_2}{2}\Big).    
\end{split}
\label{fullpnn2}
\end{equation}
Here, the CP-odd phase $\xi$ comes from $\eta$. The CP-even phases $\phi_1$ and $\phi_2$ are defined by
\begin{equation}
\phi_{1,2} \equiv \Big[ (m_n \mp \vert \mu_{n} B \vert -m_{\eta})^2 + 4 \delta^2 \Big]^{\frac{1}{2}} t. 
\end{equation}
The mixing angles $\theta_{1}$ and $\theta_{2}$ satisfy the following expression:
\begin{equation}
\tan{(2\theta_{1,2})} \equiv \frac{2\delta }{m_n \mp \vert \mu_{n} \vert B -m_{\eta}},
\label{theta12}
\end{equation}
where the magnetic moment of the neutron is given by $\mu_{n} = g_n \mu_N/2$ and $\mu_N \simeq 3.152 \times 10^{-18}$ MeV$\cdot$G$^{-1}$ is the nuclear magneton \cite{tiesinga2021codata,tiesinga2021codata2}. The parameter $\delta$ is defined by \cite{mckeen2018neutron} 
\begin{equation}
\begin{split}
\delta \equiv \frac{\lambda_{ij}\mu_{kl} \lvert \psi_q (0) \rvert^2 }{m_{\phi}^2},
\end{split}
\label{smalldelta}
\end{equation}
where $\lambda_{ij}$ is the coupling parameter between quarks and $\mu_{kl}$ is the coupling parameter between the quark and the neutral particle $\eta$. The subscripts $i$, $j$, $k$, and $l$ represent generation indices. Color antisymmetry requires that the indices $i$ and $j$ on $\lambda_{ij}$ satisfies the condition: $i\neq j$ \cite{dev2015tev}. The quark overlap parameter of the neutron takes the value: $|\psi_q (0)|^2 = 0.0144 (3)(21)$ GeV$^3$, which is calculated by lattice QCD  \cite{aoki2017improved}. Here, the numbers in the parentheses are the statistical and systematic uncertainties, respectively. We assume that the quark overlap parameter of the antineutron is equal to that of the neutron. $m_{\phi}$ is the mass of the color-multiplet scalar bosons that account for the $n$-$\eta$ mixing and can be considered as the energy scale associated with new physics.

The color-multiplet scalar bosons can mediate other effects, such as the flavor-changing neutral current (FCNC) effects
(see e.g. Refs. \cite{mohapatra2008diquark,babu2009neutrino,saha2010constraining,dorvsner2010light,giudice2011flavourful,dorvsner2011limits,barr2012observable,babu2013post,babu2013expectations,fortes2013flavor,patra2014post,sahoo2015scalar,addazi2015exotic,dev2015tev,kim2019correlation,fridell2021probing}).
Since the choice of experimental data and theoretical models is different, the phenomenological constraints on the product of the coupling parameters $|\lambda_{ij}\mu_{kl}|$ in the literature (see e.g. Refs. \cite{mohapatra2008diquark,babu2009neutrino,saha2010constraining,dorvsner2010light,giudice2011flavourful,dorvsner2011limits,barr2012observable,babu2013post,babu2013expectations,fortes2013flavor,patra2014post,sahoo2015scalar,addazi2015exotic,dev2015tev,kim2019correlation,fridell2021probing}) vary remarkably, making it difficult to compare them. If the lower bounds on the mass of the color multiplet bosons (i.e. the new physics energy scale) are restricted to the range from several TeV to several $10$ TeV that is accessible to a direct detection at the LHC or future high-energy experiments, the upper bounds on the product of the coupling parameters $|\lambda_{ij}\mu_{kl}|$ can preferably be restricted in the range from the order of $10^{-4}$ to the order of 1, roughly.

The CP asymmetry parameter $|A_{\text{CP}}|$ describes the absolute distinction between matter and antimatter. The absolute value of the CP asymmetry parameter $|A_{\text{CP}}|$ in the $n$-$\bar{n}$ oscillation process has the following maximum \cite{hao2022neutron}:
\begin{equation}
\begin{split}
\lvert A_{\text{CP}}^{\text{max}}(B=0)\rvert &=\frac{\lvert P_{n \rightarrow \bar{n}} - P_{\bar{n} \rightarrow n}\lvert}{\lvert P_{n \rightarrow \bar{n}} + P_{\bar{n} \rightarrow n}\lvert}\\
&= \frac{m_n m_{\eta} \sin^2(2\theta)}{2 m_n^2 \sin^4{\theta}+ 2m_{\eta}^2\cos^4{\theta}}.    
\end{split}
\label{a}
\end{equation}
For convenience, we have used the symbol $\lvert A_{\text{CP}}^{\text{max}}(B=0)\rvert$ to denote the maximum value of the CP asymmetry parameter, though $\lvert A_{\text{CP}}(B=0)\rvert^{\text{max}}$ may seem to be a less confusing notation.

\section{Result and Discussion}

\subsection{$n$-$\bar{n}$ Oscillation in Magnetic Fields\label{masslife}}

\begin{table*}
\caption{Comparison between the peak values of the $n$-$\bar{n}$ oscillation probability in the resonance magnetic fields and that in the absence of magnetic fields in the cases of various parameters.}
\begin{ruledtabular}
\begin{tabular}{l| l| l| l| l| l| l| l}                                         
 $|\lambda_{ij}\mu_{kl}|$ & $m_{\phi}$ (MeV) &  $\Delta m$ (MeV) & $-B_{m}$ (G) & $P_{n \rightarrow \bar{n}} (-B_{m})$ &  $B_{m}$ (G)& $P_{n \rightarrow \bar{n}} (B_{m})$ &$P_{n \rightarrow \bar{n}} (B=0)$ \\\hline           

 $1.0 \times 10^{-8}$ & $8.0 \times 10^{6}$ &  $1.0 \times 10^{-10}$  & $-1.6582 \times 10^{7}$ &  $1.27 \times 10^{-10}$  & $1.6582 \times 10^{7}$  &  $6.33 \times 10^{-11}$ & $1.03\times 10^{-18}$\\ 

 $1.0 \times 10^{-8}$ & $8.0 \times 10^{6}$ & $3.0 \times 10^{-10}$ & $-4.9745\times 10^{7}$ &  $1.41 \times 10^{-11}$ & $4.9745\times 10^{7}$ &  $7.03 \times 10^{-12}$ & $1.27 \times 10^{-20}$ \\

$1.0 \times 10^{-8}$ & $8.0 \times 10^{6}$ & $5.0 \times 10^{-10}$ &  $-8.2908 \times 10^{7}$ &  $5.06 \times 10^{-12}$ & $8.2908\times 10^{7}$ &  $2.53 \times 10^{-12}$ &  $1.64 \times 10^{-21}$ \\

 $1.0 \times 10^{-8}$ & $8.0 \times 10^{7}$ &  $1.0 \times 10^{-12}$  & $-1.6582 \times 10^{5}$ &  $1.27 \times 10^{-10}$  & $1.6582 \times 10^{5}$  &  $6.33 \times 10^{-11}$ & $1.03\times 10^{-18}$\\ 

 $1.0 \times 10^{-8}$ & $8.0 \times 10^{7}$ & $2.7 \times 10^{-12}$ & $-4.5220\times 10^{5}$ &  $1.70 \times 10^{-11}$ & $4.5220\times 10^{5}$ &  $8.51 \times 10^{-12}$ & $1.85 \times 10^{-20}$ \\

$1.0 \times 10^{-8}$ & $8.0 \times 10^{7}$ & $5.0 \times 10^{-12}$ &  $-8.2908 \times 10^{5}$ &  $5.06 \times 10^{-12}$ & $8.2908\times 10^{5}$ &  $2.53 \times 10^{-12}$ &  $1.64 \times 10^{-21}$ \\

 $1.0 \times 10^{-8}$ & $1.0 \times 10^{7}$ & $1.0 \times 10^{-10}$  &  $-1.6582 \times 10^{7}$ &  $5.18 \times 10^{-11}$  &  $1.6582 \times 10^{7}$    &  $2.59 \times 10^{-11}$& $1.72 \times 10^{-19}$   \\

 $1.0 \times 10^{-8}$ & $3.0 \times 10^{7}$ & $1.0 \times 10^{-10}$ &  $-1.6582 \times 10^{7}$ &   $6.40 \times 10^{-13}$  & $1.6582 \times 10^{7}$ &  $3.20 \times 10^{-13}$ & $2.62 \times 10^{-23}$\\

 $1.0 \times 10^{-8}$  & $1.0 \times 10^{8}$ & $1.0 \times 10^{-10}$ &  $-1.6582 \times 10^{7}$ &  $5.18 \times 10^{-15}$  & $1.6582 \times 10^{7}$ &  $2.59 \times 10^{-15}$ & $1.72 \times 10^{-27}$ \\

 $1.0 \times 10^{-6}$ & $3.0 \times 10^{7}$ & $1.0 \times 10^{-10}$  &  $-1.6582 \times 10^{7}$ &  $6.40 \times 10^{-9}$  &  $1.6582 \times 10^{7}$    &  $3.20 \times 10^{-9}$& $2.62 \times 10^{-15}$   \\ 

 $1.0 \times 10^{-7}$ & $3.0 \times 10^{7}$ & $1.0 \times 10^{-10}$  &  $-1.6582 \times 10^{7}$ &  $6.40 \times 10^{-11}$  &  $1.6582 \times 10^{7}$    &  $3.20 \times 10^{-11}$& $2.62 \times 10^{-19}$   \\

 $1.0 \times 10^{-8}$ & $3.0 \times 10^{7}$ & $1.0 \times 10^{-10}$  &  $-1.6582 \times 10^{7}$ &  $6.40 \times 10^{-13}$  &  $1.6582 \times 10^{7}$    &  $3.20 \times 10^{-13}$& $2.62 \times 10^{-23}$   \\
\end{tabular}
\end{ruledtabular}
\centering
\label{tabone}
\end{table*}

In this subsection, we focus on the probability ($P_{n \rightarrow \bar{n}}^{B \neq 0}$) of the $n$-$\bar{n}$ oscillation arising from the $n$-$\eta$ mixing in the presence of external magnetic fields. Although $\eta$ and $n^{\prime}$ are two completely different particles, both particles can lead to the $n$-$\bar{n}$ oscillation. The observability of $n$-$\bar{n}$ oscillation arising from the  $n$-$n^{\prime}$ mixing (or oscillation) in the presence of non-zero magnetic fields has been theoretically suggested in previous studies (e.g. see Refs. \cite{berezhiani2021possible,addazi2021new}).
Inspired by such studies  (e.g. see Refs. \cite{berezhiani2021possible,addazi2021new}), we explore the possible enhancement effect of the $n$-$\bar{n}$ oscillation resulting from the $n$-$\eta$ mixing by adjusting the external magnetic fields. We also evaluate the probability of the $n$-$\bar{n}$ oscillation in the cases of various parameters.

In the presence of external magnetic fields, the probability of $n$-$\bar{n}$ oscillation resulting from $n$-$\eta$ mixing in the limit $\phi_{1,2} \gg 1$ can be given by the formula (see Ref. \cite{hao2022neutron} for the derivation),
\begin{widetext}
\begin{equation}
\begin{split}
P_{n \rightarrow \bar{n}}^{B \neq 0} \simeq & \frac{1}{8} \Biggr[ \frac{4 \delta^2}{(\Delta m-|\mu_n| B)^2+ 4\delta^2} \Biggr] 
\Biggl\{\Big( \frac{m_{n} - \Delta m}{m_n} \Big)^2 \Biggr[\frac{(\Delta m -|\mu_n| B)^2 +2 \delta^2}{(\Delta m -|\mu_n| B)^2+ 4\delta^2}+ \sqrt{ \frac{(\Delta m -|\mu_n| B)^2}{(\Delta m -|\mu_n| B)^2+ 4\delta^2} } \Biggr] \\
&+ \Biggr[\frac{(\Delta m-|\mu_n| B)^2+2 \delta^2}{(\Delta m -|\mu_n| B)^2+ 4\delta^2}- \sqrt{ \frac{(\Delta m -|\mu_n| B)^2}{(\Delta m -|\mu_n| B)^2+ 4\delta^2} }\Biggr] \Biggl\} 
\Biggr[ \frac{4 \delta^2}{(\Delta m +|\mu_n| B)^2+ 4\delta^2} \Biggr]. 
\end{split}
\end{equation}
\label{pnnmagnew}
\end{widetext}
where $\Delta m \equiv m_{n} -m_{\eta}$ is the mass difference between $n$ and $\eta$.

The results of the searches for the $n$-$\bar{n}$ oscillation have been reported by various experiments \cite{baldo1994new,jones1984search,takita1986search, berger1990search,chung2002search,aharmim2017search,abe2015search,abe2021neutron}. So far, no significant evidence for the $n$-$\bar{n}$ oscillation has been found. In field-free vacuum, the lower bound on the $n$-$\bar{n}$ oscillation time reported by the ILL experiment is about $8.6 \times 10^7$ s \cite{baldo1994new}, which approximately corresponds to the oscillation probability of the order $\sim 10^{-18}$ (see e.g. Ref. \cite{berezhiani2021possible}). To facilitate comparisons, the oscillation times in  bound states reported by numerous experiments \cite{jones1984search,takita1986search, berger1990search,chung2002search,aharmim2017search,abe2015search,abe2021neutron} can be converted into oscillation probabilities using the nuclear suppression factors. The upper bounds on the oscillation probabilities can be approximately restricted in the range from the order of $10^{-20}$ to the order of $10^{-18}$ (see Refs. \cite{baldo1994new,jones1984search,takita1986search, berger1990search,chung2002search,aharmim2017search,abe2015search,abe2021neutron} for the original oscillation times, and see Ref. \cite{hao2022neutron} for the converted oscillation probabilities).

So far, the experimental searches for the $n$-$\bar{n}$ oscillation mediated by the elementary neutral particle $\eta$ have not been designed or performed. However, the experimental searches for the $n$-$\bar{n}$ oscillation mediated by the composite particle $n^{\prime}$ have been designed \cite{addazi2021new,ayres2022improved} and partially performed with respect to the $n$-$n^{\prime}$ oscillation \cite{broussard2022experimental}. Unlike the $n$-$\bar{n}$ oscillation mediated by $n^{\prime}$, the $n$-$\bar{n}$ oscillation mediated by $\eta$ has some additional advantages in the search for new physics (see e.g. Ref. \cite{hao2022neutron}) and may facilitate a better optimization of the $n$-$\bar{n}$ oscillation experiment. Nevertheless, the $n$-$\bar{n}$ oscillation mediated by $n^{\prime}$ may provide useful insights into future analysis of the $n$-$\bar{n}$ oscillation mediated by $\eta$.

\begin{figure*}[t]
\centering
\begin{subfigure}[h]{0.49\textwidth}
\centering
\includegraphics[width=\textwidth]{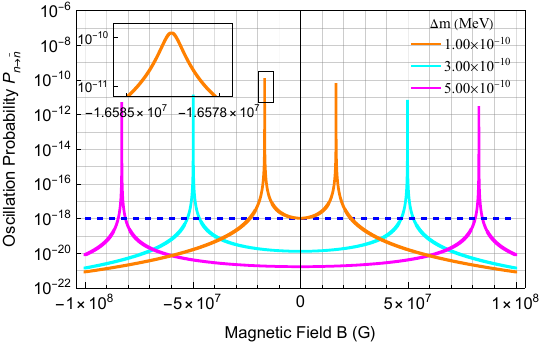}
\caption{$|\lambda_{ij}\mu_{kl}| \equiv 10^{-8}$ and $m_{\phi}\equiv 8$ TeV}
\label{fig1a}
\end{subfigure}
\hfill
\centering
\begin{subfigure}[h]{0.49\textwidth}
\centering         
\includegraphics[width=\textwidth]{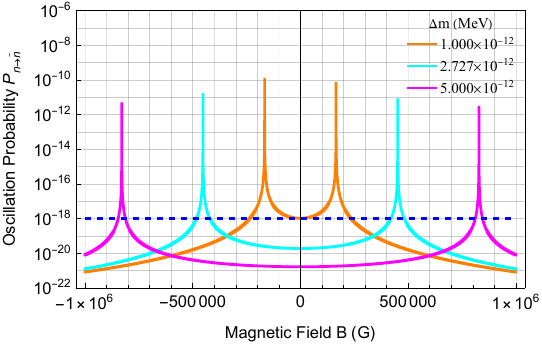}
\caption{ $|\lambda_{ij}\mu_{kl}| \equiv 10^{-8}$ and $m_{\phi}\equiv 80$ TeV}
\label{fig1b}
\end{subfigure}  
\caption{(color online) The $n$-$\bar{n}$ oscillation probability ($P_{n \rightarrow \bar{n}}^{B \neq 0}$) as a function of the applied magnetic fields B (G) in two representative cases of the parameters: $|\lambda_{ij}\mu_{kl}| \equiv 10^{-8}$ and $m_{\phi}\equiv 8$ TeV and $|\lambda_{ij}\mu_{kl}| \equiv 10^{-8}$ and $m_{\phi}\equiv 80$ TeV. The horizontal dashed line represents the upper bound on the $n$-$\bar{n}$ oscillation probability ($\sim 10^{-18}$).}
\label{fig1}
\end{figure*}

\begin{figure*}[t]
\centering
\begin{subfigure}[h]{0.49\textwidth}
\centering
\includegraphics[width=\textwidth]{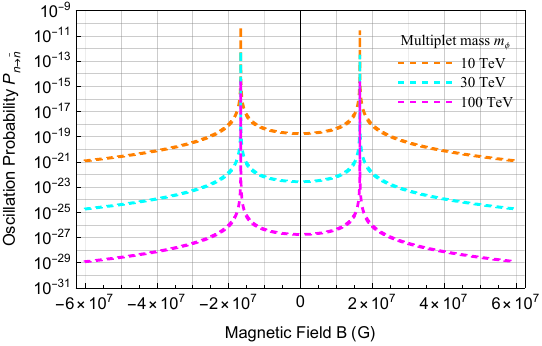}
\caption{$|\lambda_{ij}\mu_{kl}| \equiv 10^{-8}$ and $\Delta m \equiv 1.0 \times 10^{-10}$ MeV}
\label{fig2a}
\end{subfigure}
\hfill
\centering
\begin{subfigure}[h]{0.49\textwidth}
\centering         
\includegraphics[width=\textwidth]{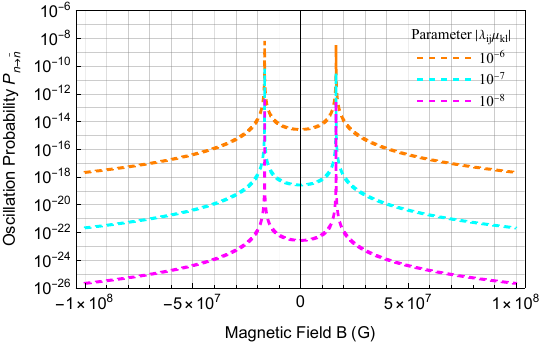}
\caption{$m_{\phi} \equiv 30$ TeV and $\Delta m \equiv 1.0 \times10^{-10}$ MeV}
\label{fig2b}
\end{subfigure}  
\caption{(color online) The $n$-$\bar{n}$ oscillation probability ($P_{n \rightarrow \bar{n}}^{B \neq 0}$) as a function of the applied magnetic fields B (G) in two representative cases of the parameters: $|\lambda_{ij}\mu_{kl}| \equiv 10^{-8}$ and $\Delta m \equiv 1.0 \times 10^{-10}$ MeV and $m_{\phi} \equiv 30$ TeV and $\Delta m \equiv 1.0 \times10^{-10}$ MeV.}
\label{fig2}
\end{figure*}

Currently, there are no exact experimental values of the coupling parameters and the masses associated with new physics beyond the SM \cite{particle2022review}. However, the numerical simulation of new physics would inevitably depend on the choice of such parameters. To analyze the effect of such parameters, we evaluate the probabilities of the $n$-$\bar{n}$ oscillation with regard to various typical values of the parameters. The corresponding results are summarized in Tab. \ref{tabone}. $P_{n \rightarrow \bar{n}} (\pm B_{m})$ represents the peak values of the oscillation probability and $\pm B_{m}$ represents the resonance values of the magnetic field, at which the peaks emerge. In the ILL experiment, the measurement was performed in magnetic fields as low as $B \lesssim 1 \times 10^{-8}$ T \cite{baldo1994new}. This value can equivalently be considered as zero magnetic field in our analysis. In order to demonstrate the enhancement effect, the probabilities of the $n$-$\bar{n}$ oscillation in the absence of the magnetic fields are also presented in Tab. \ref{tabone}. The schematic overviews of the results are presented in figures \ref{fig1}-\ref{fig2}, where a striking enhancement effect can be identified by a glimpse.

Figure \ref{fig1} shows the $n$-$\bar{n}$ oscillation probability ($P_{n \rightarrow \bar{n}}^{B \neq 0}$) resulting from the $n$-$\eta$ mixing as a function of the applied magnetic field $B$ (G) in the case of various mass differences. The results are presented for two representative cases: $|\lambda_{ij}\mu_{kl}| \equiv 10^{-8}$, $m_{\phi} \equiv 8$ TeV (Fig. \ref{fig1a}) and $|\lambda_{ij}\mu_{kl}| \equiv 10^{-8}$, $m_{\phi} \equiv 80$ TeV (Fig. \ref{fig1b}). The horizontal dashed line represents the upper bound on the $n$-$\bar{n}$ oscillation probability ($\sim 10^{-18}$), which is derived from the original $n$-$\bar{n}$ oscillation time reported by the ILL experiment \cite{baldo1994new}. In order to satisfy the bounds imposed by the experimental searches for $n$-$\bar{n}$ oscillations, the mass difference $\Delta m$ cannot be randomly chosen. In addition to this, since the resonance magnetic field, at which the peak value emerges, is determined by $\Delta m$, $\Delta m$ can be chosen based on the strongest magnetic field available in laboratories. In the search for $n$-$n^{\prime}$ oscillation, the magnitude of the applied magnetic field can be chosen as large as several Tesla (e.g. $6.6$ T \cite{broussard2022experimental}), which roughly corresponds to the mass difference on the order of $\Delta m  \sim 10^{-13}$ MeV (e.g. $\Delta m  \simeq 4 \times 10^{-13}$ MeV). Currently, the strongest steady magnetic field has been reported by the Chinese Steady High Magnetic Field Facility (SHMFF) with a value of about $45.22$ T ($4.522 \times 10^{5}$ G) \cite{zhou2023analysis}, which corresponds to $\Delta m \equiv 2.727 \times 10^{-12}$, approximately. For illustrative purposes, we choose some typical values of $\Delta m$ that are greater than the current operating limit to evaluate the oscillation probabilities in Fig. \ref{fig1a} ($\Delta m \equiv 1.0 \times 10^{-10}$, $3.0 \times 10^{-10}$, and $5.0 \times 10^{-10}$ MeV) and Fig. \ref{fig1b} ($\Delta m  \equiv 1.000 \times 10^{-12}$, $2.727 \times 10^{-12}$, and $5.000 \times 10^{-12}$ MeV). In both figures, the $n$-$\bar{n}$ oscillation probability continues to decrease as the mass difference $\Delta m$ increases.

\begin{figure*}[t]
\centering
\begin{subfigure}[h]{0.49\textwidth}
\centering
\includegraphics[width=\textwidth]{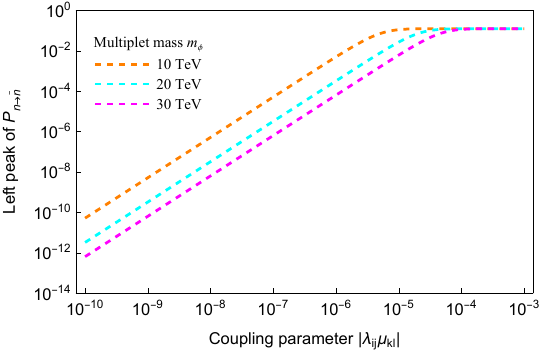}
\caption{Peak value in the negative direction}
\label{fig3a}
\end{subfigure}
\hfill
\centering
\begin{subfigure}[h]{0.49\textwidth}
\centering         
\includegraphics[width=\textwidth]{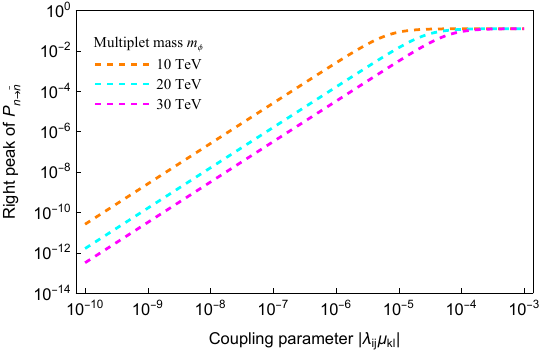}
\caption{Peak value in the positive direction}
\label{fig3b}
\end{subfigure}  
\caption{(color online) The peak values of the $n$-$\bar{n}$ oscillation probability ($P_{n \rightarrow \bar{n}}^{B \neq 0}$) as a function of the coupling parameters $|\lambda_{ij}\mu_{kl}|$ in the representative case of $\Delta m \equiv 1.0 \times 10^{-12}$ MeV.}
\label{fig3}
\end{figure*}

\begin{figure*}[t]
\centering
\begin{subfigure}[h]{0.49\textwidth}
\centering
\includegraphics[width=\textwidth]{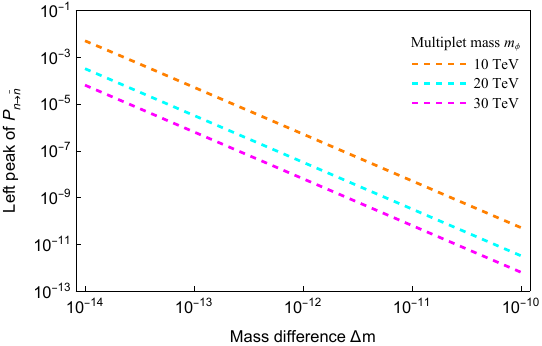}
\caption{Peak value in the negative direction}
\label{fig4a}
\end{subfigure}
\hfill
\centering
\begin{subfigure}[h]{0.49\textwidth}
\centering         
\includegraphics[width=\textwidth]{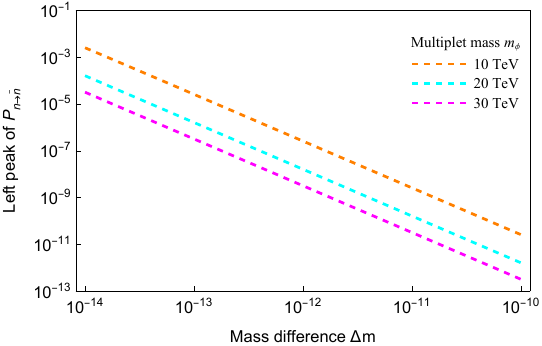}
\caption{Peak value in the positive direction}
\label{fig4b}
\end{subfigure} 
\caption{(color online) The peak values of the $n$-$\bar{n}$ oscillation probability ($P_{n \rightarrow \bar{n}}^{B \neq 0}$) as a function of the mass difference $\Delta m$ in the representative case of $|\lambda_{ij}\mu_{kl}| \equiv 10^{-8}$.}
\label{fig4}
\end{figure*}

As can be seen from Fig. \ref{fig1}, the most striking features of the curves are the two prominent peaks at certain values of the magnetic fields. One peak emerges in the positive direction of magnetic fields and the other emerges in the negative direction of magnetic fields. Our predictions are very different from the previous study \cite{berezhiani2021possible}, where there is only one peak. The peaks imply a significant enhancement effect. Such an enhancement effect can be achieved if the magnetic energy of a neutron or an antineutron in magnetic fields is equal to the mass difference between $n$ and $\eta$. Since the magnetic dipole moment of a neutron is oppositely oriented to the one of an antineutron \cite{phillips2016neutron}, two peaks could emerge at the resonant magnetic fields which differ from each other only by a minus sign. A zoomed portion of the peak area is included in figure \ref{fig1a}. By scanning the values of the magnetic field in fairly small intervals, the curves would initially rise slowly with the magnetic field. When the strength of the magnetic field exceeds a threshold value, the curves would rise rapidly. Finally, the curves peak at a certain value of the magnetic field before falling rapidly. Although similar trends can also be observed when the direction of the magnetic field is reversed ($B \rightarrow -B$), the curves in the positive direction of the magnetic field behave differently from those in the negative direction of the magnetic field. Figure \ref{fig1a} and \ref{fig1b} show remarkably that the oscillation probabilities are not symmetric with respect to reversing the direction of the applied magnetic fields. In particular, such an asymmetry would even be more significant for some specific values of the parameters.

In the model under discussion, the entire $n$-$\bar{n}$ oscillation process can be accomplished with 3 sub-processes \cite{hao2022neutron}: (1) $n$-$\eta$ oscillation, (2) $\eta$-$\bar{\eta}$ oscillation, (3) $\bar{\eta}$-$\bar{n}$ oscillation. When the direction of the magnetic field is reversed, the value of the $n$-$\eta$ oscillation probability will swap with that of the $\bar{\eta}$-$\bar{n}$ oscillation probability, suggesting that the sub-processes (1) and (3) do not account for the difference in the oscillation probabilities with respect to reversing the direction of the magnetic field. However, the $\eta$-$\bar{\eta}$ oscillation sub-process only depends on the $n$-$\eta$ mixing angle $\theta_1$ and the magnetic moment of the neutron, but does not depend on the $\bar{n}$-$\bar{\eta}$ mixing angle $\theta_2$ and the magnetic moment of the antineutron [see Eq. (\ref{theta12})]. Consequently, both the oscillation probability of the $\eta$-$\bar{\eta}$ oscillation sub-process and that of the entire $n$-$\bar{n}$ oscillation process take different values when the direction of the magnetic field is reversed. Furthermore, the CP conjugate process, namely $\bar{n}$-$n$ oscillation, can also be achieved through 3 sub-processes \cite{hao2022neutron}: (1) $\bar{n}$-$\bar{\eta}$ oscillation, (2) $\bar{\eta}$-$\eta$ oscillation, (3) $\eta$-$n$ oscillation. Unlike the $\eta$-$\bar{\eta}$ oscillation, the $\bar{\eta}$-$\eta$ oscillation only depends on the $\bar{n}$-$\bar{\eta}$ mixing angle $\theta_2$ and the magnetic moment of the antineutron and in the absence of CP-violation it behaves in the opposite way of the $\eta$-$\bar{\eta}$ oscillation sub-process. The peak value of the $n$-$\bar{n}$ oscillation probability in the negative direction of the magnetic field is approximately two times higher than that in the positive direction. On the contrary, the peak value of the $\bar{n}$-$n$ oscillation probability in the positive direction of the magnetic field is approximately two times higher than that in the negative direction.

The peak values of the $n$-$\bar{n}$ oscillation probability are approximately $8$-$10$ orders of magnitude higher than the ones in a vanishing or very small magnetic fields (e.g. $B \lesssim 10^{-8}$ T), where the ILL experiment was performed. In other words, if the applied magnetic field is adjusted properly, the $n$-$\bar{n}$ oscillation probability can be amplified by approximately $8$-$10$ orders. As an example, if the peak value emerges at the current strongest steady magnetic field ($45.22$ T), the oscillation probabilities would grow suddenly to the values of $1.70 \times 10^{-11}$ ($-B_m$) and $8.51 \times 10^{-12}$ ($B_m$), which are approximately $9$ orders of magnitude higher than the one ($1.85 \times 10^{-20}$) in a vanishing magnetic field.

Figure \ref{fig2a} shows the $n$-$\bar{n}$ oscillation probability ($P_{n \rightarrow \bar{n}}^{B \neq 0}$) as a function of the applied magnetic field $B$ in the cases of various masses of the color multiplet bosons. The results are presented in the representative case: $|\lambda_{ij}\mu_{kl}| \equiv 10^{-8}$ and $\Delta m \equiv 1.0 \times 10^{-10}$ MeV. Up to date, there has been a lack of direct experimental information on the mass of the color multiplet bosons ($m_{\phi}$, i.e. the new physics energy scale). We are only interested in the appealing scenario where the mass of the color multiplet bosons is accessible to a direct detection at the LHC or future high-energy experiments, which roughly corresponds to the range from several TeV to several $10$ TeV. For illustration, we choose some typical values of $m_{\phi}$ (i.e. $m_{\phi} \equiv 10$, $30$ and $100$ TeV) that may in general lie within the reach of direct searches, or at least not too far from the directly detectable range. Figure \ref{fig2a} suggests that the oscillation probability decreases rapidly with regard to increasing the mass of the color multiplet bosons $m_{\phi}$ and this trend is in line with our intuition.

Figure \ref{fig2b} shows the $n$-$\bar{n}$ oscillation probability ($P_{n \rightarrow \bar{n}}^{B \neq 0}$) as a function of the applied magnetic field $B$ in the cases of various coupling parameters. The results are presented in the representative case: $m_{\phi}=30$ TeV and $\Delta m=5.0 \times10^{-10}$ MeV. The curves highlight the effects of various coupling parameters on the $n$-$\bar{n}$ oscillation probability $P_{n \rightarrow \bar{n}}^{B \neq 0}$. Currently, there is no direct experimental information on the coupling parameters associated with $\eta$ either. However, the results of numerical simulations would inevitably depend on the choice of such parameters. For the purpose of illustration, we choose some typical values of $|\lambda_{ij}\mu_{kl}|$ (i.e. $|\lambda_{ij}\mu_{kl}| \equiv 10^{-8}$, $10^{-7}$ and $10^{-6}$) that are in general consistent with the FCNC constraints (i.e. $\lesssim 10^{-4}$-$10^{0}$). Figure \ref{fig2b} suggests that a decline in the $n$-$\bar{n}$ oscillation probability can be expected with regard to decreasing the coupling parameter $|\lambda_{ij}\mu_{kl}|$ and this trend meets our expectation too.

In the search for the $n$-$\bar{n}$ oscillation, the peak-detection process faces some difficulties. From theoretical aspects, the exact location of the peaks is unknown. From experimental aspects, the peaks are so narrow that it might be difficult to find them without carefully adjusting the magnetic field and choosing sufficiently large samples, especially when the homogeneity of the magnetic field is not as good as expected. Previous studies also show that some difficult experimental conditions can be further relaxed if a large and sufficiently uniform magnetic field is applied (see e.g. Ref. \cite{gudkov2020new}). Recently, technological improvements that can help reduce the magnetic disturbances from the environment have been made in the searches for the electric dipole moment of the neutron (see. e.g. Ref. \cite{ayres2022very}). Such improvements may help maintain a uniform magnetic field in the searches for the $n$-$\bar{n}$ oscillation. Since no significant signal for neutron disappearance associated with the $n$-$n^\prime$ oscillation has been found in magnetic fields below $6.6$ T \cite{broussard2022experimental}, it is natural to anticipate that the resonance values ($\pm B_{m}$) of the magnetic field might be very large.

In figures \ref{fig1}-\ref{fig2}, the curves of the $n$-$\bar{n}$ oscillation probability show downward trends in the regions that are far from the peaks. If the absolute value of the magnetic field exceeds the resonance values where the peaks emerge, the probability of the $n$-$\bar{n}$ oscillation would initially decline rapidly in the nearby regions of the peaks and then decline steadily in the distant regions of the peaks. Therefore, a large value of the magnetic field may not necessarily lead to the enhancement effect but may instead lead to a suppression effect, making the $n$-$\bar{n}$ oscillation unlikely to be observed. The magnetic fields under discussion are so strong that they are largely not reachable in laboratories with the currently available experimental techniques. However, the environments with such strong magnetic fields do exist in our universe. For example, highly magnetized neutron stars, namely magnetars, possess remarkably powerful magnetic fields, which are typically as strong as $10^{13}$-$10^{15}$ G near the surface and expected to be even stronger in the vicinity of the core \cite{mereghetti2015magnetars}. Although carrying out experiments in such strong magnetic fields is largely infeasible either at present or in the near future, our work offers possibilities and hope for future explorations. Nevertheless, our simulations on the $n$-$\bar{n}$ oscillation in strong magnetic fields can not only help develop the experimental setup for further detection, but also can help interpret the results more correctly after the experiments.

\begin{figure}[b] 
\centering
\includegraphics[scale=1.0,width=0.99\linewidth]{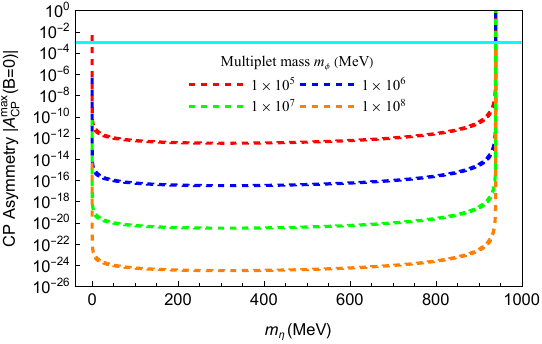}
\caption{(color online) The maximum value of the CP asymmetry parameter $\lvert A_{\text{CP}}^{\text{max}}($B=0$)\rvert$ as a function of $m_{\eta}$ in the representative cases of various masses of the color multiplet bosons ($\lvert\lambda_{ij}\mu_{kl} \lvert \equiv 10^{-1}$).}
\label{fig5}
\end{figure}

\begin{figure}[b] 
\centering
\includegraphics[scale=1.0,width=0.99\linewidth]{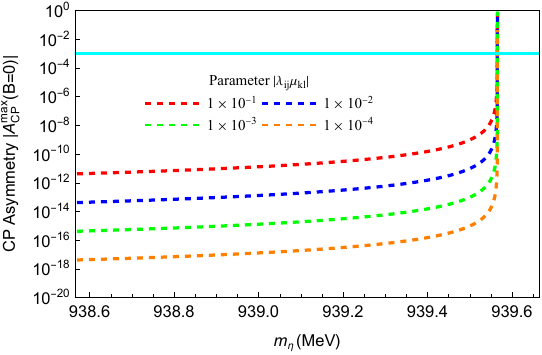}
\caption{(color online) The maximum value of the CP asymmetry parameter $\lvert A_{\text{CP}}^{\text{max}}($B=0$)\rvert$ as a function of $m_{\eta}$ for the representative cases of various coupling parameters ($m_{\phi} \equiv 1$ TeV).}
\label{fig6}
\end{figure}

The peak values of the $n$-$\bar{n}$ oscillation probability ($P_{n \rightarrow \bar{n}}^{B \neq 0}$) also depend on the choice of the parameters, such as $|\lambda_{ij}\mu_{kl}|$, $m_{\eta}$ (or equivalently $\Delta m$), etc. The effects of different parameters are correlated with each other. Moreover, no direct experimental information on such parameters can be found. We try to analyze the sensitivity of the peak values with respect to one of such parameters, with the others held constant values. The results are summarized in Fig. \ref{fig3} and \ref{fig4}. 
Fig. \ref{fig3} shows the peak values of the $n$-$\bar{n}$ oscillation probability ($P_{n \rightarrow \bar{n}}^{B \neq 0}$) as a function of the coupling parameters $|\lambda_{ij}\mu_{kl}|$. The curves are obtained by evaluating the peak values at some typical constant values of $m_{\phi}$ ($10$, $20$, and $30$ TeV) and $\Delta m$ ($1.0 \times 10^{-12}$ MeV). Fig. \ref{fig3a} (Fig. \ref{fig3b}) corresponds to the peak values in the negative (positive) direction of magnetic fields. The curves rise steadily in the lower range of $|\lambda_{ij}\mu_{kl}|$ and then level off in the upper range of $|\lambda_{ij}\mu_{kl}|$. 
Fig. \ref{fig4} shows the peak values of the $n$-$\bar{n}$ oscillation probability ($P_{n \rightarrow \bar{n}}^{B \neq 0}$) as a function of the mass difference $\Delta m$. The curves are obtained by evaluating the peak values at some typical constant values of $m_{\phi}$ ($10$, $20$, and $30$ TeV) and $|\lambda_{ij}\mu_{kl}|$ ($1.0 \times 10^{-8}$). Fig. \ref{fig4a} (Fig. \ref{fig4b}) corresponds to the peak values in the negative (positive) direction of magnetic fields. The curves fall steadily within the range from the order of $10^{-14}$ to the order of $10^{-10}$. In both figures, although the results are sensitive to the choice of the parameters, they are still many orders of magnitude higher than the current experimental limits. Therefore, our work offers possibilities and hope for future experimental explorations though carrying out the experiments in such strong magnetic fields is infeasible with the currently available experimental techniques.

\subsection{CP-violation  \label{sec4}}

In this subsection, we explore the possibility of observing CP-violation in the $n$-$\bar{n}$ oscillation process resulting from the $n$-$\eta$ mixing. We choose the typical size of the CP asymmetry ($|A_{\text{CP}}|$) in the meson systems as the benchmark value of CP-violation and estimate its constraints on $m_{\eta}$ in the $n$-$\bar{n}$ oscillation mediated by $\eta$. We also analyze the effects of the parameters by evaluating $|A^{\text{max}}_{\text{CP}}(B=0)|$ in some representative cases of the coupling parameters and the masses of the color multiplet boson.

CP violation has been confirmed in various meson systems, such as the neutral $K$ meson \cite{christenson1964evidence}, $B$ meson \cite{aubert2001observation,abe2001observation}, and $D$ meson \cite{aaij2019observation}. The reported magnitudes of the CP asymmetry in the meson systems depend on the specific processes and roughly lie within the range from the order of $10^{-3}$ to the order of $10^{-2}$ (see e.g. Ref. \cite{particle2022review}). To be conservative, we choose $\lvert A_{\text{CP}}^{\text{max}}(B=0)\rvert \equiv 1 \times 10^{-3}$ as the representative benchmark value of the CP asymmetry (i.e. the typical size of the CP asymmetry in the meson systems) in our analysis.

Figure \ref{fig5} shows the maximum value of the CP asymmetry parameter $\lvert A_{\text{CP}}^{\text{max}}($B=0$)\rvert$ as a function of $m_{\eta}$ in the cases of various masses of the color multiplet bosons. The results are presented in the representative case of $|\lambda_{ij}\mu_{kl}| \equiv 10^{-1}$. Specifically, the dashed curves represent the derived upper bounds on $\lvert A_{\text{CP}}^{\text{max}}($B=0$)\rvert$ in the cases of $m_{\phi} \equiv 0.1$, $1$, $10$, $100$ TeV, which, in general, are located in the directly detectable range of the LHC or future high-energy experiments. Again, the horizontal solid line represents the typical size of the CP asymmetry in the meson systems. The derived upper bounds on $\lvert A_{\text{CP}}^{\text{max}}($B=0$)\rvert$ decrease with the mass of the color multiplet bosons. The most striking feature of the curves is that they concave up and have a U-shape over the entire allowed range of $m_\eta$. Figure \ref{fig5} also shows that the bounds on the CP asymmetry have the lowest values in the middle ranges of $m_\eta$ and have peak values in both the lower and upper ranges. The magnitude of the CP asymmetry rises rapidly when $m_{\eta}$ is approaching both $m_n$ and $0$. However, according to subsection \ref{masslife}, if $m_{\eta}$ is very small, $\Delta m$ would be so large that the $n$-$\bar{n}$ oscillation probability would be highly suppressed and thus might not lie within the detectable regions. Therefore, in order to have sizable CP-violating and $\mathcal{B}$-violating effects, the value of $m_{\eta}$ should be restricted in the vicinity of $m_n$.

Figure \ref{fig6} shows the maximum value of the CP asymmetry parameter $\lvert A_{\text{CP}}^{\text{max}}($B=0$)\rvert$ as a function of $m_{\eta}$ in the cases of various coupling parameters. The results are presented in the representative case of $m_{\phi} \equiv 1$ TeV. Specifically, the dashed curves represent the derived upper bounds on $\lvert A_{\text{CP}}^{\text{max}}($B=0$)\rvert$ in the cases of $|\lambda_{ij}\mu_{kl}| \equiv 10^{-4}$, $10^{-3}$, $10^{-2}$, $10^{-1}$, which in general correspond to the highest possible values of the coupling parameters allowed by the FCNC constraints. The curves initially rise slowly in the lower ranges of $m_{\eta}$ but then rise rapidly in the upper ranges of $m_{\eta}$. For the sake of comparison, we include the typical value of the CP-violation in the meson systems here, though the size of CP violation in the meson systems does not necessarily coincide with that in the $n$-$\bar{n}$ oscillation. The horizontal solid line represents the typical size of the CP asymmetry in the meson systems. The derived upper bounds on $\lvert A_{\text{CP}}^{\text{max}}($B=0$)\rvert$ increase as the coupling parameters increase. Figure \ref{fig5} suggests that unless the coupling parameters are sufficiently large, there might not be sizable CP-violating effects in the $n$-$\bar{n}$ oscillation. However, in subsection \ref{masslife}, we have assumed that the mass of the color multiplet bosons is accessible for direct detection at the LHC or future high-energy experiments, and in this case, the coupling parameters cannot be too large.

\section{Summary and Conclusion}

In this paper, we have briefly reviewed the model that accounts for the $n$-$\bar{n}$ oscillation mediated by $\eta$. We have also reviewed the mathematical description of the $n$-$\bar{n}$ oscillation and its connection with CP-violation. After that, we have focused on the impact of magnetic fields on the $n$-$\bar{n}$ oscillation probability. In addition to this, we have explored the possible enhancement effect of the $n$-$\bar{n}$ oscillation by adjusting the magnitude of the magnetic fields. Finally, based on the experimental results of CP-violation in the meson systems, we have discussed the observability of CP-violation in the $n$-$\bar{n}$ oscillation process.

We have evaluated the $n$-$\bar{n}$ oscillation probability ($P_{n \rightarrow \bar{n}}^{B \neq 0}$) for some representative cases of the parameters and analyzed the impact of various parameters on the observability of the $n$-$\bar{n}$ oscillation.  We have also explored the possible enhancement effect of the $n$-$\bar{n}$ oscillation by adjusting the applied magnetic fields. To satisfy the constraints imposed by the experimental searches for $n$-$\bar{n}$ oscillations, $\Delta m$ cannot be randomly chosen. For illustrative purposes, we chose some typical values of $\Delta m$. We have shown that increasing the mass difference $\Delta m$ tends to reduce the $n$-$\bar{n}$ oscillation probability. The resonance magnetic field, at which the peak value of the $n$-$\bar{n}$ oscillation probability emerges, is determined by $\Delta m$. The most striking features of the $n$-$\bar{n}$ oscillation probability are the two prominent peaks at certain values of the magnetic fields. The peak values of the $n$-$\bar{n}$ oscillation probability are approximately $8$-$10$ orders higher than the ones in the absence of magnetic fields. In other words, if the applied magnetic field is adjusted properly, the $n$-$\bar{n}$ oscillation probability can be amplified by approximately $8$-$10$ orders, offering new opportunities for observing the $n$-$\bar{n}$ oscillation. We have also pointed out that the oscillation probabilities are not symmetric with respect to flipping the direction of the applied magnetic fields. In particular, such an asymmetry would even be more significant for some specific values of the parameters.

We have also explored the possibility of observing CP-violation in the $n$-$\bar{n}$ oscillation process resulting from $n$-$\eta$ mixing. We have chosen the typical size of the CP asymmetry in the meson systems as the benchmark value of CP-violation and estimated its constraint on $m_{\eta}$ in the $n$-$\bar{n}$ oscillation mediated by $\eta$. We have analyzed how the coupling parameters ($|\lambda_{ij}\mu_{kl}|$) and the masses of color multiplet bosons ($m_{\phi}$) affect the value of $\lvert A_{\text{CP}}^{\text{max}}($B=0$)\rvert$. The derived upper bounds on $\lvert A_{\text{CP}}^{\text{max}}($B=0$)\rvert$ initially rise slowly in the lower ranges of $m_{\eta}$ but then rise rapidly in the upper ranges of $m_{\eta}$. Furthermore, the derived upper bounds on $\lvert A_{\text{CP}}^{\text{max}}($B=0$)\rvert$ increase with the coupling parameters, but decrease with the mass of the color multiplet bosons. We have shown that unless the coupling parameters are sufficiently large, there might not be sizable CP-violating effects in the  $n$-$\bar{n}$ oscillation. The most striking feature of the curves is that they concave up and have a U-shape over the entire allowed range of $m_\eta$. We have also pointed out that the bounds on the CP asymmetry have the lowest values in the middle ranges of $m_\eta$ and have peaks in both the lower and upper ranges. The magnitude of the CP asymmetry rises rapidly when $m_{\eta}$ is approaching both $m_n$ and $0$. However, if $m_{\eta}$ is very small, $\Delta m$ would be so large that the $n$-$\bar{n}$ oscillation probability might be highly suppressed and thus not lie within the detectable regions. To summarize, in order to have sizable CP-violating and $\mathcal{B}$-violating effects, $m_{\eta}$ should lie within the vicinity of $m_n$.

One of the previous similar studies shows that the probability of the $n$-$\bar{n}$ oscillation can also be enhanced by many orders \cite{berezhiani2021possible}. We consider our findings are novel due to the following reasons. To begin with, in the formalism of the SM, the phenomena of particle mixing usually either refer to the mixing between two elementary particles or refer to the mixing between two composite particles. The mixing between the composite and the elementary particles is relatively rare to see in the literature. In the previous similar study \cite{berezhiani2021possible}, the $n$-$\bar{n}$ oscillation is induced by the mixing between the neutron (composite particle) and mirror-neutron (composite particle). Therefore, we consider the $n$-$\bar{n}$ oscillation induced by the mixing between the neutron (composite particle) and the neutral particle $\eta$ (elementary particle) in the presence of magnetic fields is a novel and interesting subject. Moreover, previous study shows that although the $n$-$\bar{n}$ oscillation process can be considered as a potential probe for CP-violation \cite{berezhiani2015neutronantineutron}, such a process may not necessarily lead to observable CP-violating effects, because the phase associated with CP-violation in this process can be absorbed into the definition of the neutron field \cite{fujikawa2015neutronantineutron}. In our work, we show explicitly that the $n$-$\bar{n}$ oscillation accompanied by CP-violation could lead to observable effects in the presence of magnetic fields. Finally, in our work, the most striking feature of the $n$-$\bar{n}$ oscillation is that it has two prominent peaks at certain values of magnetic fields. One peak emerges in the positive direction of magnetic fields and the other emerges in the negative direction of magnetic fields. Our predictions are very different from the previous study \cite{berezhiani2021possible}, where there is only one peak. If two peaks of the $n$-$\bar{n}$ oscillation were observed with respect to flipping the direction of the applied magnetic fields, our predictions would be justified.

\section*{Acknowledgement}
This work is supported by the National Natural Science Foundation of China (Grant No. 12104187), Macao Young Scholars Program (No. AM2021001), Jiangsu Provincial Double-Innovation Doctor Program (Grant No. JSSCBS20210940), and the Startup Funding of Jiangsu University (No. 4111710002). Y. H. and K. S. N. contributed equally to this work.


\bibliography{nnbar}

\begin{thebibliography}{92}%
\makeatletter
\providecommand \@ifxundefined [1]{%
 \@ifx{#1\undefined}
}%
\providecommand \@ifnum [1]{%
 \ifnum #1\expandafter \@firstoftwo
 \else \expandafter \@secondoftwo
 \fi
}%
\providecommand \@ifx [1]{%
 \ifx #1\expandafter \@firstoftwo
 \else \expandafter \@secondoftwo
 \fi
}%
\providecommand \natexlab [1]{#1}%
\providecommand \enquote  [1]{``#1''}%
\providecommand \bibnamefont  [1]{#1}%
\providecommand \bibfnamefont [1]{#1}%
\providecommand \citenamefont [1]{#1}%
\providecommand \href@noop [0]{\@secondoftwo}%
\providecommand \href [0]{\begingroup \@sanitize@url \@href}%
\providecommand \@href[1]{\@@startlink{#1}\@@href}%
\providecommand \@@href[1]{\endgroup#1\@@endlink}%
\providecommand \@sanitize@url [0]{\catcode `\\12\catcode `\$12\catcode `\&12\catcode `\#12\catcode `\^12\catcode `\_12\catcode `\%12\relax}%
\providecommand \@@startlink[1]{}%
\providecommand \@@endlink[0]{}%
\providecommand \url  [0]{\begingroup\@sanitize@url \@url }%
\providecommand \@url [1]{\endgroup\@href {#1}{\urlprefix }}%
\providecommand \urlprefix  [0]{URL }%
\providecommand \Eprint [0]{\href }%
\providecommand \doibase [0]{http://dx.doi.org/}%
\providecommand \selectlanguage [0]{\@gobble}%
\providecommand \bibinfo  [0]{\@secondoftwo}%
\providecommand \bibfield  [0]{\@secondoftwo}%
\providecommand \translation [1]{[#1]}%
\providecommand \BibitemOpen [0]{}%
\providecommand \bibitemStop [0]{}%
\providecommand \bibitemNoStop [0]{.\EOS\space}%
\providecommand \EOS [0]{\spacefactor3000\relax}%
\providecommand \BibitemShut  [1]{\csname bibitem#1\endcsname}%
\let\auto@bib@innerbib\@empty
\bibitem [{\citenamefont {Snow}\ \emph {et~al.}(2022)\citenamefont {Snow}, \citenamefont {Haddock},\ and\ \citenamefont {Heacock}}]{snow2022searches}%
  \BibitemOpen
  \bibfield  {author} {\bibinfo {author} {\bibfnamefont {W.~M.}\ \bibnamefont {Snow}}, \bibinfo {author} {\bibfnamefont {C.}~\bibnamefont {Haddock}}, \ and\ \bibinfo {author} {\bibfnamefont {B.}~\bibnamefont {Heacock}},\ }\href {\doibase 10.3390/sym14010010} {\bibfield  {journal} {\bibinfo  {journal} {Symmetry}\ }\textbf {\bibinfo {volume} {14}},\ \bibinfo {pages} {10} (\bibinfo {year} {2022})}\BibitemShut {NoStop}%
\bibitem [{\citenamefont {Alarcon}\ \emph {et~al.}(2023)\citenamefont {Alarcon}, \citenamefont {Aleksandrova}, \citenamefont {Bae{\ss}ler}, \citenamefont {Beck}, \citenamefont {Bhattacharya}, \citenamefont {Blatnik}, \citenamefont {Bowles}, \citenamefont {Bowman}, \citenamefont {Brewington}, \citenamefont {Broussard} \emph {et~al.}}]{alarcon2023fundamental}%
  \BibitemOpen
  \bibfield  {author} {\bibinfo {author} {\bibfnamefont {R.}~\bibnamefont {Alarcon}}, \bibinfo {author} {\bibfnamefont {A.}~\bibnamefont {Aleksandrova}}, \bibinfo {author} {\bibfnamefont {S.}~\bibnamefont {Bae{\ss}ler}}, \bibinfo {author} {\bibfnamefont {D.~H.}\ \bibnamefont {Beck}}, \bibinfo {author} {\bibfnamefont {T.}~\bibnamefont {Bhattacharya}}, \bibinfo {author} {\bibfnamefont {M.}~\bibnamefont {Blatnik}}, \bibinfo {author} {\bibfnamefont {T.~J.}\ \bibnamefont {Bowles}}, \bibinfo {author} {\bibfnamefont {J.~D.}\ \bibnamefont {Bowman}}, \bibinfo {author} {\bibfnamefont {J.}~\bibnamefont {Brewington}}, \bibinfo {author} {\bibfnamefont {L.~J.}\ \bibnamefont {Broussard}},  \emph {et~al.},\ }\href {https://arxiv.org/abs/2308.09059} {\bibfield  {journal} {\bibinfo  {journal} {arXiv:2308.09059}\ } (\bibinfo {year} {2023})}\BibitemShut {NoStop}%
\bibitem [{\citenamefont {Dubbers}\ and\ \citenamefont {Schmidt}(2011)}]{dubbers2011neutron}%
  \BibitemOpen
  \bibfield  {author} {\bibinfo {author} {\bibfnamefont {D.}~\bibnamefont {Dubbers}}\ and\ \bibinfo {author} {\bibfnamefont {M.~G.}\ \bibnamefont {Schmidt}},\ }\href {\doibase 10.1103/RevModPhys.83.1111} {\bibfield  {journal} {\bibinfo  {journal} {Rev. Mod. Phys.}\ }\textbf {\bibinfo {volume} {83}},\ \bibinfo {pages} {1111} (\bibinfo {year} {2011})}\BibitemShut {NoStop}%
\bibitem [{\citenamefont {Yue}\ \emph {et~al.}(2013)\citenamefont {Yue}, \citenamefont {Dewey}, \citenamefont {Gilliam}, \citenamefont {Greene}, \citenamefont {Laptev}, \citenamefont {Nico}, \citenamefont {Snow},\ and\ \citenamefont {Wietfeldt}}]{yue2013improved}%
  \BibitemOpen
  \bibfield  {author} {\bibinfo {author} {\bibfnamefont {A.~T.}\ \bibnamefont {Yue}}, \bibinfo {author} {\bibfnamefont {M.~S.}\ \bibnamefont {Dewey}}, \bibinfo {author} {\bibfnamefont {D.~M.}\ \bibnamefont {Gilliam}}, \bibinfo {author} {\bibfnamefont {G.~L.}\ \bibnamefont {Greene}}, \bibinfo {author} {\bibfnamefont {A.~B.}\ \bibnamefont {Laptev}}, \bibinfo {author} {\bibfnamefont {J.~S.}\ \bibnamefont {Nico}}, \bibinfo {author} {\bibfnamefont {W.~M.}\ \bibnamefont {Snow}}, \ and\ \bibinfo {author} {\bibfnamefont {F.~E.}\ \bibnamefont {Wietfeldt}},\ }\href {\doibase 10.1103/PhysRevLett.111.222501} {\bibfield  {journal} {\bibinfo  {journal} {Phys. Rev. Lett.}\ }\textbf {\bibinfo {volume} {111}},\ \bibinfo {pages} {222501} (\bibinfo {year} {2013})}\BibitemShut {NoStop}%
\bibitem [{\citenamefont {Gonzalez}\ \emph {et~al.}(2021)\citenamefont {Gonzalez}, \citenamefont {Fries}, \citenamefont {Cude-Woods}, \citenamefont {Bailey}, \citenamefont {Blatnik}, \citenamefont {Broussard}, \citenamefont {Callahan}, \citenamefont {Choi}, \citenamefont {Clayton}, \citenamefont {Currie} \emph {et~al.}}]{gonzalez2021improved}%
  \BibitemOpen
  \bibfield  {author} {\bibinfo {author} {\bibfnamefont {F.~M.}\ \bibnamefont {Gonzalez}}, \bibinfo {author} {\bibfnamefont {E.~M.}\ \bibnamefont {Fries}}, \bibinfo {author} {\bibfnamefont {C.}~\bibnamefont {Cude-Woods}}, \bibinfo {author} {\bibfnamefont {T.}~\bibnamefont {Bailey}}, \bibinfo {author} {\bibfnamefont {M.}~\bibnamefont {Blatnik}}, \bibinfo {author} {\bibfnamefont {L.~J.}\ \bibnamefont {Broussard}}, \bibinfo {author} {\bibfnamefont {N.~B.}\ \bibnamefont {Callahan}}, \bibinfo {author} {\bibfnamefont {J.~H.}\ \bibnamefont {Choi}}, \bibinfo {author} {\bibfnamefont {S.~M.}\ \bibnamefont {Clayton}}, \bibinfo {author} {\bibfnamefont {S.~A.}\ \bibnamefont {Currie}},  \emph {et~al.},\ }\href {\doibase 10.1103/PhysRevLett.127.162501} {\bibfield  {journal} {\bibinfo  {journal} {Phys. Rev. Lett.}\ }\textbf {\bibinfo {volume} {127}},\ \bibinfo {pages} {162501} (\bibinfo {year} {2021})}\BibitemShut {NoStop}%
\bibitem [{\citenamefont {Workman}\ \emph {et~al.}(2022)\citenamefont {Workman}, \citenamefont {Burkert}, \citenamefont {Crede}, \citenamefont {Klempt}, \citenamefont {Thoma}, \citenamefont {Tiator}, \citenamefont {Agashe}, \citenamefont {Aielli}, \citenamefont {Allanach} \emph {et~al.}}]{particle2022review}%
  \BibitemOpen
  \bibfield  {author} {\bibinfo {author} {\bibfnamefont {R.~L.}\ \bibnamefont {Workman}}, \bibinfo {author} {\bibfnamefont {V.~D.}\ \bibnamefont {Burkert}}, \bibinfo {author} {\bibfnamefont {V.}~\bibnamefont {Crede}}, \bibinfo {author} {\bibfnamefont {E.}~\bibnamefont {Klempt}}, \bibinfo {author} {\bibfnamefont {U.}~\bibnamefont {Thoma}}, \bibinfo {author} {\bibfnamefont {L.}~\bibnamefont {Tiator}}, \bibinfo {author} {\bibfnamefont {K.}~\bibnamefont {Agashe}}, \bibinfo {author} {\bibfnamefont {G.}~\bibnamefont {Aielli}}, \bibinfo {author} {\bibfnamefont {B.~C.}\ \bibnamefont {Allanach}},  \emph {et~al.} (\bibinfo {collaboration} {Particle Data Group}),\ }\href {\doibase 10.1093/ptep/ptac097} {\bibfield  {journal} {\bibinfo  {journal} {Prog. Theor. Exp. Phys.}\ }\textbf {\bibinfo {volume} {2022}},\ \bibinfo {pages} {083C01} (\bibinfo {year} {2022})}\BibitemShut {NoStop}%
\bibitem [{\citenamefont {Tan}(2023)}]{tan2023neutron}%
  \BibitemOpen
  \bibfield  {author} {\bibinfo {author} {\bibfnamefont {W.}~\bibnamefont {Tan}},\ }\href {\doibase 10.3390/universe9040180} {\bibfield  {journal} {\bibinfo  {journal} {Universe}\ }\textbf {\bibinfo {volume} {9}},\ \bibinfo {pages} {180} (\bibinfo {year} {2023})}\BibitemShut {NoStop}%
\bibitem [{\citenamefont {Mohapatra}(2009)}]{mohapatra2009neutron}%
  \BibitemOpen
  \bibfield  {author} {\bibinfo {author} {\bibfnamefont {R.~N.}\ \bibnamefont {Mohapatra}},\ }\href {\doibase 10.1088/0954-3899/36/10/104006} {\bibfield  {journal} {\bibinfo  {journal} {J. Phys. G: Nucl. Part. Phys.}\ }\textbf {\bibinfo {volume} {36}},\ \bibinfo {pages} {104006} (\bibinfo {year} {2009})}\BibitemShut {NoStop}%
\bibitem [{\citenamefont {Phillips~II}\ \emph {et~al.}(2016)\citenamefont {Phillips~II}, \citenamefont {Snow}, \citenamefont {Babu}, \citenamefont {Banerjee}, \citenamefont {Baxter}, \citenamefont {Berezhiani}, \citenamefont {Bergevin}, \citenamefont {Bhattacharya}, \citenamefont {Brooijmans}, \citenamefont {Castellanos} \emph {et~al.}}]{phillips2016neutron}%
  \BibitemOpen
  \bibfield  {author} {\bibinfo {author} {\bibfnamefont {D.~G.}\ \bibnamefont {Phillips~II}}, \bibinfo {author} {\bibfnamefont {W.~M.}\ \bibnamefont {Snow}}, \bibinfo {author} {\bibfnamefont {K.}~\bibnamefont {Babu}}, \bibinfo {author} {\bibfnamefont {S.}~\bibnamefont {Banerjee}}, \bibinfo {author} {\bibfnamefont {D.~V.}\ \bibnamefont {Baxter}}, \bibinfo {author} {\bibfnamefont {Z.}~\bibnamefont {Berezhiani}}, \bibinfo {author} {\bibfnamefont {M.}~\bibnamefont {Bergevin}}, \bibinfo {author} {\bibfnamefont {S.}~\bibnamefont {Bhattacharya}}, \bibinfo {author} {\bibfnamefont {G.}~\bibnamefont {Brooijmans}}, \bibinfo {author} {\bibfnamefont {L.}~\bibnamefont {Castellanos}},  \emph {et~al.},\ }\href {\doibase 10.1016/j.physrep.2015.11.001} {\bibfield  {journal} {\bibinfo  {journal} {Phys. Rep.}\ }\textbf {\bibinfo {volume} {612}},\ \bibinfo {pages} {1} (\bibinfo {year} {2016})}\BibitemShut {NoStop}%
\bibitem [{\citenamefont {Baldo-Ceolin}\ \emph {et~al.}(1994)\citenamefont {Baldo-Ceolin}, \citenamefont {Benetti}, \citenamefont {Bitter}, \citenamefont {Bobisut}, \citenamefont {Calligarich}, \citenamefont {Dolfini}, \citenamefont {Dubbers}, \citenamefont {El-Muzeini}, \citenamefont {Genoni}, \citenamefont {Gibin} \emph {et~al.}}]{baldo1994new}%
  \BibitemOpen
  \bibfield  {author} {\bibinfo {author} {\bibfnamefont {M.}~\bibnamefont {Baldo-Ceolin}}, \bibinfo {author} {\bibfnamefont {P.}~\bibnamefont {Benetti}}, \bibinfo {author} {\bibfnamefont {T.}~\bibnamefont {Bitter}}, \bibinfo {author} {\bibfnamefont {F.}~\bibnamefont {Bobisut}}, \bibinfo {author} {\bibfnamefont {E.}~\bibnamefont {Calligarich}}, \bibinfo {author} {\bibfnamefont {R.}~\bibnamefont {Dolfini}}, \bibinfo {author} {\bibfnamefont {D.}~\bibnamefont {Dubbers}}, \bibinfo {author} {\bibfnamefont {P.}~\bibnamefont {El-Muzeini}}, \bibinfo {author} {\bibfnamefont {M.}~\bibnamefont {Genoni}}, \bibinfo {author} {\bibfnamefont {D.}~\bibnamefont {Gibin}},  \emph {et~al.},\ }\href {\doibase 10.1007/BF01580321} {\bibfield  {journal} {\bibinfo  {journal} {Z. Phys. C}\ }\textbf {\bibinfo {volume} {63}},\ \bibinfo {pages} {409} (\bibinfo {year} {1994})}\BibitemShut {NoStop}%
\bibitem [{\citenamefont {Jones}\ \emph {et~al.}(1984)\citenamefont {Jones}, \citenamefont {Bionta},\ and\ \citenamefont {Blewitt}}]{jones1984search}%
  \BibitemOpen
  \bibfield  {author} {\bibinfo {author} {\bibfnamefont {T.~W.}\ \bibnamefont {Jones}}, \bibinfo {author} {\bibfnamefont {R.~M.}\ \bibnamefont {Bionta}}, \ and\ \bibinfo {author} {\bibfnamefont {G.}~\bibnamefont {Blewitt}},\ }\href {\doibase 10.1103/PhysRevLett.52.720} {\bibfield  {journal} {\bibinfo  {journal} {Phys. Rev. Lett.}\ }\textbf {\bibinfo {volume} {52}},\ \bibinfo {pages} {720} (\bibinfo {year} {1984})}\BibitemShut {NoStop}%
\bibitem [{\citenamefont {Takita}\ \emph {et~al.}(1986)\citenamefont {Takita}, \citenamefont {Arisaka}, \citenamefont {Kajita}, \citenamefont {Kifune}, \citenamefont {Koshiba}, \citenamefont {Miyano}, \citenamefont {Nakahata}, \citenamefont {Oyama}, \citenamefont {Sato}, \citenamefont {Suda} \emph {et~al.}}]{takita1986search}%
  \BibitemOpen
  \bibfield  {author} {\bibinfo {author} {\bibfnamefont {M.}~\bibnamefont {Takita}}, \bibinfo {author} {\bibfnamefont {K.}~\bibnamefont {Arisaka}}, \bibinfo {author} {\bibfnamefont {T.}~\bibnamefont {Kajita}}, \bibinfo {author} {\bibfnamefont {T.}~\bibnamefont {Kifune}}, \bibinfo {author} {\bibfnamefont {M.}~\bibnamefont {Koshiba}}, \bibinfo {author} {\bibfnamefont {K.}~\bibnamefont {Miyano}}, \bibinfo {author} {\bibfnamefont {M.}~\bibnamefont {Nakahata}}, \bibinfo {author} {\bibfnamefont {Y.}~\bibnamefont {Oyama}}, \bibinfo {author} {\bibfnamefont {N.}~\bibnamefont {Sato}}, \bibinfo {author} {\bibfnamefont {T.}~\bibnamefont {Suda}},  \emph {et~al.},\ }\href {\doibase 10.1103/PhysRevD.34.902} {\bibfield  {journal} {\bibinfo  {journal} {Phys. Rev. D}\ }\textbf {\bibinfo {volume} {34}},\ \bibinfo {pages} {902} (\bibinfo {year} {1986})}\BibitemShut {NoStop}%
\bibitem [{\citenamefont {Berger}\ \emph {et~al.}(1990)\citenamefont {Berger}, \citenamefont {Fr{\"o}hlich}, \citenamefont {M{\"o}nch}, \citenamefont {Nisius}, \citenamefont {Raupach}, \citenamefont {Schleper}, \citenamefont {Benadjal}, \citenamefont {Blum}, \citenamefont {Bourdarios}, \citenamefont {Dudelzak} \emph {et~al.}}]{berger1990search}%
  \BibitemOpen
  \bibfield  {author} {\bibinfo {author} {\bibfnamefont {C.}~\bibnamefont {Berger}}, \bibinfo {author} {\bibfnamefont {M.}~\bibnamefont {Fr{\"o}hlich}}, \bibinfo {author} {\bibfnamefont {H.}~\bibnamefont {M{\"o}nch}}, \bibinfo {author} {\bibfnamefont {R.}~\bibnamefont {Nisius}}, \bibinfo {author} {\bibfnamefont {F.}~\bibnamefont {Raupach}}, \bibinfo {author} {\bibfnamefont {P.}~\bibnamefont {Schleper}}, \bibinfo {author} {\bibfnamefont {Y.}~\bibnamefont {Benadjal}}, \bibinfo {author} {\bibfnamefont {D.}~\bibnamefont {Blum}}, \bibinfo {author} {\bibfnamefont {C.}~\bibnamefont {Bourdarios}}, \bibinfo {author} {\bibfnamefont {B.}~\bibnamefont {Dudelzak}},  \emph {et~al.},\ }\href {\doibase 10.1016/0370-2693(90)90441-8} {\bibfield  {journal} {\bibinfo  {journal} {Phys. Lett. B}\ }\textbf {\bibinfo {volume} {240}},\ \bibinfo {pages} {237} (\bibinfo {year} {1990})}\BibitemShut {NoStop}%
\bibitem [{\citenamefont {Chung}\ \emph {et~al.}(2002)\citenamefont {Chung}, \citenamefont {Allison}, \citenamefont {Alner}, \citenamefont {Ayres}, \citenamefont {Barrett}, \citenamefont {Border}, \citenamefont {Cobb}, \citenamefont {Courant}, \citenamefont {Demuth}, \citenamefont {Fields} \emph {et~al.}}]{chung2002search}%
  \BibitemOpen
  \bibfield  {author} {\bibinfo {author} {\bibfnamefont {J.}~\bibnamefont {Chung}}, \bibinfo {author} {\bibfnamefont {W.}~\bibnamefont {Allison}}, \bibinfo {author} {\bibfnamefont {G.~J.}\ \bibnamefont {Alner}}, \bibinfo {author} {\bibfnamefont {D.~S.}\ \bibnamefont {Ayres}}, \bibinfo {author} {\bibfnamefont {W.~L.}\ \bibnamefont {Barrett}}, \bibinfo {author} {\bibfnamefont {P.~M.}\ \bibnamefont {Border}}, \bibinfo {author} {\bibfnamefont {J.~H.}\ \bibnamefont {Cobb}}, \bibinfo {author} {\bibfnamefont {H.}~\bibnamefont {Courant}}, \bibinfo {author} {\bibfnamefont {D.~M.}\ \bibnamefont {Demuth}}, \bibinfo {author} {\bibfnamefont {T.~H.}\ \bibnamefont {Fields}},  \emph {et~al.},\ }\href {\doibase 10.1103/PhysRevD.66.032004} {\bibfield  {journal} {\bibinfo  {journal} {Phys. Rev. D}\ }\textbf {\bibinfo {volume} {66}},\ \bibinfo {pages} {032004} (\bibinfo {year} {2002})}\BibitemShut {NoStop}%
\bibitem [{\citenamefont {Aharmim}\ \emph {et~al.}(2017)\citenamefont {Aharmim}, \citenamefont {Ahmed}, \citenamefont {Anthony}, \citenamefont {Barros}, \citenamefont {Beier}, \citenamefont {Bellerive}, \citenamefont {Beltran}, \citenamefont {Bergevin}, \citenamefont {Biller}, \citenamefont {Boudjemline} \emph {et~al.}}]{aharmim2017search}%
  \BibitemOpen
  \bibfield  {author} {\bibinfo {author} {\bibfnamefont {B.}~\bibnamefont {Aharmim}}, \bibinfo {author} {\bibfnamefont {S.~N.}\ \bibnamefont {Ahmed}}, \bibinfo {author} {\bibfnamefont {A.~E.}\ \bibnamefont {Anthony}}, \bibinfo {author} {\bibfnamefont {N.}~\bibnamefont {Barros}}, \bibinfo {author} {\bibfnamefont {E.~W.}\ \bibnamefont {Beier}}, \bibinfo {author} {\bibfnamefont {A.}~\bibnamefont {Bellerive}}, \bibinfo {author} {\bibfnamefont {B.}~\bibnamefont {Beltran}}, \bibinfo {author} {\bibfnamefont {M.}~\bibnamefont {Bergevin}}, \bibinfo {author} {\bibfnamefont {S.~D.}\ \bibnamefont {Biller}}, \bibinfo {author} {\bibfnamefont {K.}~\bibnamefont {Boudjemline}},  \emph {et~al.},\ }\href {\doibase 10.1103/PhysRevD.96.092005} {\bibfield  {journal} {\bibinfo  {journal} {Phys. Rev. D}\ }\textbf {\bibinfo {volume} {96}},\ \bibinfo {pages} {092005} (\bibinfo {year} {2017})}\BibitemShut {NoStop}%
\bibitem [{\citenamefont {Abe}\ \emph {et~al.}(2015)\citenamefont {Abe}, \citenamefont {Hayato}, \citenamefont {Iida}, \citenamefont {Ishihara}, \citenamefont {Kameda}, \citenamefont {Koshio}, \citenamefont {Minamino}, \citenamefont {Mitsuda}, \citenamefont {Miura}, \citenamefont {Moriyama} \emph {et~al.}}]{abe2015search}%
  \BibitemOpen
  \bibfield  {author} {\bibinfo {author} {\bibfnamefont {K.}~\bibnamefont {Abe}}, \bibinfo {author} {\bibfnamefont {Y.}~\bibnamefont {Hayato}}, \bibinfo {author} {\bibfnamefont {T.}~\bibnamefont {Iida}}, \bibinfo {author} {\bibfnamefont {K.}~\bibnamefont {Ishihara}}, \bibinfo {author} {\bibfnamefont {J.}~\bibnamefont {Kameda}}, \bibinfo {author} {\bibfnamefont {Y.}~\bibnamefont {Koshio}}, \bibinfo {author} {\bibfnamefont {A.}~\bibnamefont {Minamino}}, \bibinfo {author} {\bibfnamefont {C.}~\bibnamefont {Mitsuda}}, \bibinfo {author} {\bibfnamefont {M.}~\bibnamefont {Miura}}, \bibinfo {author} {\bibfnamefont {S.}~\bibnamefont {Moriyama}},  \emph {et~al.},\ }\href {\doibase 10.1103/PhysRevD.91.072006} {\bibfield  {journal} {\bibinfo  {journal} {Phys. Rev. D}\ }\textbf {\bibinfo {volume} {91}},\ \bibinfo {pages} {072006} (\bibinfo {year} {2015})}\BibitemShut {NoStop}%
\bibitem [{\citenamefont {Abe}\ \emph {et~al.}(2021)\citenamefont {Abe}, \citenamefont {Bronner}, \citenamefont {Hayato}, \citenamefont {Ikeda}, \citenamefont {Imaizumi}, \citenamefont {Ito}, \citenamefont {Kameda}, \citenamefont {Kataoka}, \citenamefont {Miura}, \citenamefont {Moriyama} \emph {et~al.}}]{abe2021neutron}%
  \BibitemOpen
  \bibfield  {author} {\bibinfo {author} {\bibfnamefont {K.}~\bibnamefont {Abe}}, \bibinfo {author} {\bibfnamefont {C.}~\bibnamefont {Bronner}}, \bibinfo {author} {\bibfnamefont {Y.}~\bibnamefont {Hayato}}, \bibinfo {author} {\bibfnamefont {M.}~\bibnamefont {Ikeda}}, \bibinfo {author} {\bibfnamefont {S.}~\bibnamefont {Imaizumi}}, \bibinfo {author} {\bibfnamefont {H.}~\bibnamefont {Ito}}, \bibinfo {author} {\bibfnamefont {J.}~\bibnamefont {Kameda}}, \bibinfo {author} {\bibfnamefont {Y.}~\bibnamefont {Kataoka}}, \bibinfo {author} {\bibfnamefont {M.}~\bibnamefont {Miura}}, \bibinfo {author} {\bibfnamefont {S.}~\bibnamefont {Moriyama}},  \emph {et~al.},\ }\href {\doibase 10.1103/PhysRevD.103.012008} {\bibfield  {journal} {\bibinfo  {journal} {Phys. Rev. D}\ }\textbf {\bibinfo {volume} {103}},\ \bibinfo {pages} {012008} (\bibinfo {year} {2021})}\BibitemShut {NoStop}%
\bibitem [{\citenamefont {Sandars}(1980)}]{sandars1980neutron}%
  \BibitemOpen
  \bibfield  {author} {\bibinfo {author} {\bibfnamefont {P.~G.~H.}\ \bibnamefont {Sandars}},\ }\href {\doibase 10.1088/0305-4616/6/10/001} {\bibfield  {journal} {\bibinfo  {journal} {J. Phys. G: Nucl. Phys.}\ }\textbf {\bibinfo {volume} {6}},\ \bibinfo {pages} {L161} (\bibinfo {year} {1980})}\BibitemShut {NoStop}%
\bibitem [{\citenamefont {Alberico}\ \emph {et~al.}(1982)\citenamefont {Alberico}, \citenamefont {Bottino},\ and\ \citenamefont {Molinari}}]{alberico1982new}%
  \BibitemOpen
  \bibfield  {author} {\bibinfo {author} {\bibfnamefont {W.~M.}\ \bibnamefont {Alberico}}, \bibinfo {author} {\bibfnamefont {A.}~\bibnamefont {Bottino}}, \ and\ \bibinfo {author} {\bibfnamefont {A.}~\bibnamefont {Molinari}},\ }\href {\doibase 10.1016/0370-2693(82)90493-2} {\bibfield  {journal} {\bibinfo  {journal} {Phys. Lett. B}\ }\textbf {\bibinfo {volume} {114}},\ \bibinfo {pages} {266} (\bibinfo {year} {1982})}\BibitemShut {NoStop}%
\bibitem [{\citenamefont {Dover}\ \emph {et~al.}(1983)\citenamefont {Dover}, \citenamefont {Gal},\ and\ \citenamefont {Richard}}]{dover1983neutron}%
  \BibitemOpen
  \bibfield  {author} {\bibinfo {author} {\bibfnamefont {C.~B.}\ \bibnamefont {Dover}}, \bibinfo {author} {\bibfnamefont {A.}~\bibnamefont {Gal}}, \ and\ \bibinfo {author} {\bibfnamefont {J.~M.}\ \bibnamefont {Richard}},\ }\href {\doibase 10.1103/PhysRevD.27.1090} {\bibfield  {journal} {\bibinfo  {journal} {Phys. Rev. D}\ }\textbf {\bibinfo {volume} {27}},\ \bibinfo {pages} {1090} (\bibinfo {year} {1983})}\BibitemShut {NoStop}%
\bibitem [{\citenamefont {Addazi}\ \emph {et~al.}(2021)\citenamefont {Addazi}, \citenamefont {Anderson}, \citenamefont {Ansell}, \citenamefont {Babu}, \citenamefont {Barrow}, \citenamefont {Baxter}, \citenamefont {Bentley}, \citenamefont {Berezhiani}, \citenamefont {Bevilacqua}, \citenamefont {Biondi} \emph {et~al.}}]{addazi2021new}%
  \BibitemOpen
  \bibfield  {author} {\bibinfo {author} {\bibfnamefont {A.}~\bibnamefont {Addazi}}, \bibinfo {author} {\bibfnamefont {K.}~\bibnamefont {Anderson}}, \bibinfo {author} {\bibfnamefont {S.}~\bibnamefont {Ansell}}, \bibinfo {author} {\bibfnamefont {K.~S.}\ \bibnamefont {Babu}}, \bibinfo {author} {\bibfnamefont {J.~L.}\ \bibnamefont {Barrow}}, \bibinfo {author} {\bibfnamefont {D.~V.}\ \bibnamefont {Baxter}}, \bibinfo {author} {\bibfnamefont {P.~M.}\ \bibnamefont {Bentley}}, \bibinfo {author} {\bibfnamefont {Z.}~\bibnamefont {Berezhiani}}, \bibinfo {author} {\bibfnamefont {R.}~\bibnamefont {Bevilacqua}}, \bibinfo {author} {\bibfnamefont {R.}~\bibnamefont {Biondi}},  \emph {et~al.},\ }\href {\doibase 10.1088/1361-6471/abf429} {\bibfield  {journal} {\bibinfo  {journal} {J. Phys. G: Nucl. Part. Phys.}\ }\textbf {\bibinfo {volume} {48}},\ \bibinfo {pages} {070501} (\bibinfo {year} {2021})}\BibitemShut {NoStop}%
\bibitem [{\citenamefont {Berezhiani}\ and\ \citenamefont {Bento}(2006)}]{berezhiani2006neutron}%
  \BibitemOpen
  \bibfield  {author} {\bibinfo {author} {\bibfnamefont {Z.}~\bibnamefont {Berezhiani}}\ and\ \bibinfo {author} {\bibfnamefont {L.}~\bibnamefont {Bento}},\ }\href {\doibase 10.1103/PhysRevLett.96.081801} {\bibfield  {journal} {\bibinfo  {journal} {Phys. Rev. Lett.}\ }\textbf {\bibinfo {volume} {96}},\ \bibinfo {pages} {081801} (\bibinfo {year} {2006})}\BibitemShut {NoStop}%
\bibitem [{\citenamefont {Berezhiani}(2009)}]{berezhiani2009more}%
  \BibitemOpen
  \bibfield  {author} {\bibinfo {author} {\bibfnamefont {Z.}~\bibnamefont {Berezhiani}},\ }\href {\doibase 10.1140/epjc/s10052-009-1165-1} {\bibfield  {journal} {\bibinfo  {journal} {Eur. Phys. J. C}\ }\textbf {\bibinfo {volume} {64}},\ \bibinfo {pages} {421} (\bibinfo {year} {2009})}\BibitemShut {NoStop}%
\bibitem [{\citenamefont {Berezhiani}\ and\ \citenamefont {Nesti}(2012)}]{berezhiani2012magnetic}%
  \BibitemOpen
  \bibfield  {author} {\bibinfo {author} {\bibfnamefont {Z.}~\bibnamefont {Berezhiani}}\ and\ \bibinfo {author} {\bibfnamefont {F.}~\bibnamefont {Nesti}},\ }\href {\doibase 10.1140/epjc/s10052-012-1974-5} {\bibfield  {journal} {\bibinfo  {journal} {Eur. Phys. J. C}\ }\textbf {\bibinfo {volume} {72}},\ \bibinfo {pages} {1974} (\bibinfo {year} {2012})}\BibitemShut {NoStop}%
\bibitem [{\citenamefont {Berezhiani}(2019)}]{berezhiani2019neutron2}%
  \BibitemOpen
  \bibfield  {author} {\bibinfo {author} {\bibfnamefont {Z.}~\bibnamefont {Berezhiani}},\ }\href {\doibase 10.1140/epjc/s10052-019-6995-x} {\bibfield  {journal} {\bibinfo  {journal} {Eur. Phys. J. C}\ }\textbf {\bibinfo {volume} {79}},\ \bibinfo {pages} {484} (\bibinfo {year} {2019})}\BibitemShut {NoStop}%
\bibitem [{\citenamefont {Berezhiani}\ \emph {et~al.}(2021)\citenamefont {Berezhiani}, \citenamefont {Biondi}, \citenamefont {Mannarelli},\ and\ \citenamefont {Tonelli}}]{berezhiani2021neutron}%
  \BibitemOpen
  \bibfield  {author} {\bibinfo {author} {\bibfnamefont {Z.}~\bibnamefont {Berezhiani}}, \bibinfo {author} {\bibfnamefont {R.}~\bibnamefont {Biondi}}, \bibinfo {author} {\bibfnamefont {M.}~\bibnamefont {Mannarelli}}, \ and\ \bibinfo {author} {\bibfnamefont {F.}~\bibnamefont {Tonelli}},\ }\href {\doibase 10.1140/epjc/s10052-021-09806-1} {\bibfield  {journal} {\bibinfo  {journal} {Eur. Phys. J. C}\ }\textbf {\bibinfo {volume} {81}},\ \bibinfo {pages} {1036} (\bibinfo {year} {2021})}\BibitemShut {NoStop}%
\bibitem [{\citenamefont {Berezhiani}\ \emph {et~al.}(2019)\citenamefont {Berezhiani}, \citenamefont {Biondi}, \citenamefont {Kamyshkov},\ and\ \citenamefont {Varriano}}]{berezhiani2019on}%
  \BibitemOpen
  \bibfield  {author} {\bibinfo {author} {\bibfnamefont {Z.}~\bibnamefont {Berezhiani}}, \bibinfo {author} {\bibfnamefont {R.}~\bibnamefont {Biondi}}, \bibinfo {author} {\bibfnamefont {Y.}~\bibnamefont {Kamyshkov}}, \ and\ \bibinfo {author} {\bibfnamefont {L.}~\bibnamefont {Varriano}},\ }\href {\doibase 10.3390/physics1020021} {\bibfield  {journal} {\bibinfo  {journal} {Phys.}\ }\textbf {\bibinfo {volume} {1}},\ \bibinfo {pages} {271} (\bibinfo {year} {2019})}\BibitemShut {NoStop}%
\bibitem [{\citenamefont {Babu}\ and\ \citenamefont {Mohapatra}(2022)}]{babu2022theoretical}%
  \BibitemOpen
  \bibfield  {author} {\bibinfo {author} {\bibfnamefont {K.~S.}\ \bibnamefont {Babu}}\ and\ \bibinfo {author} {\bibfnamefont {R.~N.}\ \bibnamefont {Mohapatra}},\ }\href {\doibase 10.3390/sym14040731} {\bibfield  {journal} {\bibinfo  {journal} {Symm.}\ }\textbf {\bibinfo {volume} {14}},\ \bibinfo {pages} {731} (\bibinfo {year} {2022})}\BibitemShut {NoStop}%
\bibitem [{\citenamefont {Pokotilovski}(2006)}]{pokotilovski2006experimental}%
  \BibitemOpen
  \bibfield  {author} {\bibinfo {author} {\bibfnamefont {Y.~N.}\ \bibnamefont {Pokotilovski}},\ }\href {\doibase 10.1016/j.physletb.2006.06.005} {\bibfield  {journal} {\bibinfo  {journal} {Phys. Lett. B}\ }\textbf {\bibinfo {volume} {639}},\ \bibinfo {pages} {214} (\bibinfo {year} {2006})}\BibitemShut {NoStop}%
\bibitem [{\citenamefont {Berezhiani}\ \emph {et~al.}(2017)\citenamefont {Berezhiani}, \citenamefont {Frost}, \citenamefont {Kamyshkov}, \citenamefont {Rybolt},\ and\ \citenamefont {Varriano}}]{berezhiani2017neutron}%
  \BibitemOpen
  \bibfield  {author} {\bibinfo {author} {\bibfnamefont {Z.}~\bibnamefont {Berezhiani}}, \bibinfo {author} {\bibfnamefont {M.}~\bibnamefont {Frost}}, \bibinfo {author} {\bibfnamefont {Y.}~\bibnamefont {Kamyshkov}}, \bibinfo {author} {\bibfnamefont {B.}~\bibnamefont {Rybolt}}, \ and\ \bibinfo {author} {\bibfnamefont {L.}~\bibnamefont {Varriano}},\ }\href {\doibase 10.1103/PhysRevD.96.035039} {\bibfield  {journal} {\bibinfo  {journal} {Phys. Rev. D}\ }\textbf {\bibinfo {volume} {96}},\ \bibinfo {pages} {035039} (\bibinfo {year} {2017})}\BibitemShut {NoStop}%
\bibitem [{\citenamefont {Kamyshkov}\ \emph {et~al.}(2022)\citenamefont {Kamyshkov}, \citenamefont {Ternullo}, \citenamefont {Varriano},\ and\ \citenamefont {Berezhiani}}]{kamyshkov2022neutron}%
  \BibitemOpen
  \bibfield  {author} {\bibinfo {author} {\bibfnamefont {Y.}~\bibnamefont {Kamyshkov}}, \bibinfo {author} {\bibfnamefont {J.}~\bibnamefont {Ternullo}}, \bibinfo {author} {\bibfnamefont {L.}~\bibnamefont {Varriano}}, \ and\ \bibinfo {author} {\bibfnamefont {Z.}~\bibnamefont {Berezhiani}},\ }\href {\doibase 10.3390/sym14020230} {\bibfield  {journal} {\bibinfo  {journal} {Symm.}\ }\textbf {\bibinfo {volume} {14}},\ \bibinfo {pages} {230} (\bibinfo {year} {2022})}\BibitemShut {NoStop}%
\bibitem [{\citenamefont {Mohapatra}\ \emph {et~al.}(2005)\citenamefont {Mohapatra}, \citenamefont {Nasri},\ and\ \citenamefont {Nussinov}}]{mohapatra2005some}%
  \BibitemOpen
  \bibfield  {author} {\bibinfo {author} {\bibfnamefont {R.~N.}\ \bibnamefont {Mohapatra}}, \bibinfo {author} {\bibfnamefont {S.}~\bibnamefont {Nasri}}, \ and\ \bibinfo {author} {\bibfnamefont {S.}~\bibnamefont {Nussinov}},\ }\href {\doibase 10.1016/j.physletb.2005.08.101} {\bibfield  {journal} {\bibinfo  {journal} {Phys. Lett. B}\ }\textbf {\bibinfo {volume} {627}},\ \bibinfo {pages} {124} (\bibinfo {year} {2005})}\BibitemShut {NoStop}%
\bibitem [{\citenamefont {Berezhiani}(2016)}]{berezhiani2016neutron}%
  \BibitemOpen
  \bibfield  {author} {\bibinfo {author} {\bibfnamefont {Z.}~\bibnamefont {Berezhiani}},\ }\href {\doibase 10.1140/epjc/s10052-016-4564-0} {\bibfield  {journal} {\bibinfo  {journal} {Eur. Phys. J. C}\ }\textbf {\bibinfo {volume} {76}},\ \bibinfo {pages} {705} (\bibinfo {year} {2016})}\BibitemShut {NoStop}%
\bibitem [{\citenamefont {Berezhiani}(2021)}]{berezhiani2021possible}%
  \BibitemOpen
  \bibfield  {author} {\bibinfo {author} {\bibfnamefont {Z.}~\bibnamefont {Berezhiani}},\ }\href {\doibase 10.1140/epjc/s10052-020-08824-9} {\bibfield  {journal} {\bibinfo  {journal} {Eur. Phys. J. C}\ }\textbf {\bibinfo {volume} {81}},\ \bibinfo {pages} {33} (\bibinfo {year} {2021})}\BibitemShut {NoStop}%
\bibitem [{\citenamefont {Ayres}\ \emph {et~al.}(2022{\natexlab{a}})\citenamefont {Ayres}, \citenamefont {Berezhiani}, \citenamefont {Biondi}, \citenamefont {Bison}, \citenamefont {Bodek}, \citenamefont {Bondar}, \citenamefont {Chiu}, \citenamefont {Daum}, \citenamefont {Dinani}, \citenamefont {Doorenbos} \emph {et~al.}}]{ayres2022improved}%
  \BibitemOpen
  \bibfield  {author} {\bibinfo {author} {\bibfnamefont {N.~J.}\ \bibnamefont {Ayres}}, \bibinfo {author} {\bibfnamefont {Z.}~\bibnamefont {Berezhiani}}, \bibinfo {author} {\bibfnamefont {R.}~\bibnamefont {Biondi}}, \bibinfo {author} {\bibfnamefont {G.}~\bibnamefont {Bison}}, \bibinfo {author} {\bibfnamefont {K.}~\bibnamefont {Bodek}}, \bibinfo {author} {\bibfnamefont {V.}~\bibnamefont {Bondar}}, \bibinfo {author} {\bibfnamefont {P.-J.}\ \bibnamefont {Chiu}}, \bibinfo {author} {\bibfnamefont {M.}~\bibnamefont {Daum}}, \bibinfo {author} {\bibfnamefont {R.~T.}\ \bibnamefont {Dinani}}, \bibinfo {author} {\bibfnamefont {C.~B.}\ \bibnamefont {Doorenbos}},  \emph {et~al.},\ }\href {\doibase 10.3390/sym14030503} {\bibfield  {journal} {\bibinfo  {journal} {Symm.}\ }\textbf {\bibinfo {volume} {14}},\ \bibinfo {pages} {503} (\bibinfo {year} {2022}{\natexlab{a}})}\BibitemShut {NoStop}%
\bibitem [{\citenamefont {Broussard}\ \emph {et~al.}(2022)\citenamefont {Broussard}, \citenamefont {Barrow}, \citenamefont {DeBeer-Schmitt}, \citenamefont {Dennis}, \citenamefont {Fitzsimmons}, \citenamefont {Frost}, \citenamefont {Gilbert}, \citenamefont {Gonzalez}, \citenamefont {Heilbronn}, \citenamefont {Iverson} \emph {et~al.}}]{broussard2022experimental}%
  \BibitemOpen
  \bibfield  {author} {\bibinfo {author} {\bibfnamefont {L.~J.}\ \bibnamefont {Broussard}}, \bibinfo {author} {\bibfnamefont {J.~L.}\ \bibnamefont {Barrow}}, \bibinfo {author} {\bibfnamefont {L.}~\bibnamefont {DeBeer-Schmitt}}, \bibinfo {author} {\bibfnamefont {T.}~\bibnamefont {Dennis}}, \bibinfo {author} {\bibfnamefont {M.~R.}\ \bibnamefont {Fitzsimmons}}, \bibinfo {author} {\bibfnamefont {M.~J.}\ \bibnamefont {Frost}}, \bibinfo {author} {\bibfnamefont {C.~E.}\ \bibnamefont {Gilbert}}, \bibinfo {author} {\bibfnamefont {F.~M.}\ \bibnamefont {Gonzalez}}, \bibinfo {author} {\bibfnamefont {L.}~\bibnamefont {Heilbronn}}, \bibinfo {author} {\bibfnamefont {E.~B.}\ \bibnamefont {Iverson}},  \emph {et~al.},\ }\href {\doibase 10.1103/PhysRevLett.128.212503} {\bibfield  {journal} {\bibinfo  {journal} {Phys. Rev. Lett.}\ }\textbf {\bibinfo {volume} {128}},\ \bibinfo {pages} {212503} (\bibinfo {year} {2022})}\BibitemShut {NoStop}%
\bibitem [{\citenamefont {Acharya}\ \emph {et~al.}(2023)\citenamefont {Acharya}, \citenamefont {Adams}, \citenamefont {Aleksandrova}, \citenamefont {Alfonso}, \citenamefont {An}, \citenamefont {Bae{\ss}ler}, \citenamefont {Balantekin}, \citenamefont {Barbeau}, \citenamefont {Bellini}, \citenamefont {Bellini} \emph {et~al.}}]{acharya2023fundamental}%
  \BibitemOpen
  \bibfield  {author} {\bibinfo {author} {\bibfnamefont {B.}~\bibnamefont {Acharya}}, \bibinfo {author} {\bibfnamefont {C.}~\bibnamefont {Adams}}, \bibinfo {author} {\bibfnamefont {A.~A.}\ \bibnamefont {Aleksandrova}}, \bibinfo {author} {\bibfnamefont {K.}~\bibnamefont {Alfonso}}, \bibinfo {author} {\bibfnamefont {P.}~\bibnamefont {An}}, \bibinfo {author} {\bibfnamefont {S.}~\bibnamefont {Bae{\ss}ler}}, \bibinfo {author} {\bibfnamefont {A.~B.}\ \bibnamefont {Balantekin}}, \bibinfo {author} {\bibfnamefont {P.~S.}\ \bibnamefont {Barbeau}}, \bibinfo {author} {\bibfnamefont {F.}~\bibnamefont {Bellini}}, \bibinfo {author} {\bibfnamefont {V.}~\bibnamefont {Bellini}},  \emph {et~al.},\ }\href {https://arxiv.org/abs/2304.03451} {\bibfield  {journal} {\bibinfo  {journal} {arXiv:2304.03451}\ } (\bibinfo {year} {2023})}\BibitemShut {NoStop}%
\bibitem [{\citenamefont {Perez}\ \emph {et~al.}(2022)\citenamefont {Perez}, \citenamefont {Pocar}, \citenamefont {Babu}, \citenamefont {Broussard}, \citenamefont {Cirigliano}, \citenamefont {Gardner}, \citenamefont {Heeck}, \citenamefont {Kearns}, \citenamefont {Long}, \citenamefont {Raby} \emph {et~al.}}]{perez2022baryon}%
  \BibitemOpen
  \bibfield  {author} {\bibinfo {author} {\bibfnamefont {P.~F.}\ \bibnamefont {Perez}}, \bibinfo {author} {\bibfnamefont {A.}~\bibnamefont {Pocar}}, \bibinfo {author} {\bibfnamefont {K.~S.}\ \bibnamefont {Babu}}, \bibinfo {author} {\bibfnamefont {L.~J.}\ \bibnamefont {Broussard}}, \bibinfo {author} {\bibfnamefont {V.}~\bibnamefont {Cirigliano}}, \bibinfo {author} {\bibfnamefont {S.}~\bibnamefont {Gardner}}, \bibinfo {author} {\bibfnamefont {J.}~\bibnamefont {Heeck}}, \bibinfo {author} {\bibfnamefont {E.}~\bibnamefont {Kearns}}, \bibinfo {author} {\bibfnamefont {A.~J.}\ \bibnamefont {Long}}, \bibinfo {author} {\bibfnamefont {S.}~\bibnamefont {Raby}},  \emph {et~al.},\ }\href {https://arxiv.org/abs/2208.00010} {\bibfield  {journal} {\bibinfo  {journal} {arXiv:2208.00010}\ } (\bibinfo {year} {2022})}\BibitemShut {NoStop}%
\bibitem [{\citenamefont {Hao}\ and\ \citenamefont {Ni}(2022)}]{hao2022neutron}%
  \BibitemOpen
  \bibfield  {author} {\bibinfo {author} {\bibfnamefont {Y.}~\bibnamefont {Hao}}\ and\ \bibinfo {author} {\bibfnamefont {D.}~\bibnamefont {Ni}},\ }\href {\doibase 10.1103/PhysRevD.106.115028} {\bibfield  {journal} {\bibinfo  {journal} {Phys. Rev. D}\ }\textbf {\bibinfo {volume} {106}},\ \bibinfo {pages} {115028} (\bibinfo {year} {2022})}\BibitemShut {NoStop}%
\bibitem [{\citenamefont {Branco}\ \emph {et~al.}(1999)\citenamefont {Branco}, \citenamefont {Lavoura},\ and\ \citenamefont {Silva}}]{branco1999cp}%
  \BibitemOpen
  \bibfield  {author} {\bibinfo {author} {\bibfnamefont {G.~C.}\ \bibnamefont {Branco}}, \bibinfo {author} {\bibfnamefont {L.}~\bibnamefont {Lavoura}}, \ and\ \bibinfo {author} {\bibfnamefont {J.~P.}\ \bibnamefont {Silva}},\ }\href {https://global.oup.com/academic/product/cp-violation-9780198503996?cc=us&lang=en&#} {\emph {\bibinfo {title} {CP violation}}}\ (\bibinfo  {publisher} {Oxford University Press},\ \bibinfo {address} {Oxford},\ \bibinfo {year} {1999})\BibitemShut {NoStop}%
\bibitem [{\citenamefont {Sozzi}(2007)}]{sozzi2007discrete}%
  \BibitemOpen
  \bibfield  {author} {\bibinfo {author} {\bibfnamefont {M.}~\bibnamefont {Sozzi}},\ }\href {\doibase 10.1093/acprof:oso/9780199296668.001.0001} {\emph {\bibinfo {title} {Discrete symmetries and CP violation: From experiment to theory}}}\ (\bibinfo  {publisher} {Oxford University Press},\ \bibinfo {address} {Oxford},\ \bibinfo {year} {2007})\BibitemShut {NoStop}%
\bibitem [{\citenamefont {Bigi}\ and\ \citenamefont {Sanda}(2009)}]{bigi2009cp}%
  \BibitemOpen
  \bibfield  {author} {\bibinfo {author} {\bibfnamefont {I.~I.}\ \bibnamefont {Bigi}}\ and\ \bibinfo {author} {\bibfnamefont {A.~I.}\ \bibnamefont {Sanda}},\ }\href {\doibase 10.1017/CBO9780511581014} {\emph {\bibinfo {title} {CP violation}}}\ (\bibinfo  {publisher} {Cambridge University Press},\ \bibinfo {address} {Cambridge},\ \bibinfo {year} {2009})\BibitemShut {NoStop}%
\bibitem [{\citenamefont {Christenson}\ \emph {et~al.}(1964)\citenamefont {Christenson}, \citenamefont {Cronin}, \citenamefont {Fitch},\ and\ \citenamefont {Turlay}}]{christenson1964evidence}%
  \BibitemOpen
  \bibfield  {author} {\bibinfo {author} {\bibfnamefont {J.~H.}\ \bibnamefont {Christenson}}, \bibinfo {author} {\bibfnamefont {J.~W.}\ \bibnamefont {Cronin}}, \bibinfo {author} {\bibfnamefont {V.~L.}\ \bibnamefont {Fitch}}, \ and\ \bibinfo {author} {\bibfnamefont {R.}~\bibnamefont {Turlay}},\ }\href {\doibase 10.1103/PhysRevLett.13.138} {\bibfield  {journal} {\bibinfo  {journal} {Phys. Rev. Lett.}\ }\textbf {\bibinfo {volume} {13}},\ \bibinfo {pages} {138} (\bibinfo {year} {1964})}\BibitemShut {NoStop}%
\bibitem [{\citenamefont {Bigi}\ and\ \citenamefont {Sanada}(1981)}]{bigi1981notes}%
  \BibitemOpen
  \bibfield  {author} {\bibinfo {author} {\bibfnamefont {I.~I.}\ \bibnamefont {Bigi}}\ and\ \bibinfo {author} {\bibfnamefont {A.~I.}\ \bibnamefont {Sanada}},\ }\href {\doibase 10.1016/0550-3213(81)90519-8} {\bibfield  {journal} {\bibinfo  {journal} {Nucl. Phys. B}\ }\textbf {\bibinfo {volume} {193}},\ \bibinfo {pages} {85} (\bibinfo {year} {1981})}\BibitemShut {NoStop}%
\bibitem [{\citenamefont {Aubert}\ \emph {et~al.}(2001)\citenamefont {Aubert}, \citenamefont {Boutigny}, \citenamefont {Gaillard}, \citenamefont {Hicheur}, \citenamefont {Karyotakis}, \citenamefont {Lees}, \citenamefont {Robbe}, \citenamefont {Tisserand}, \citenamefont {Palano}, \citenamefont {Chen} \emph {et~al.}}]{aubert2001observation}%
  \BibitemOpen
  \bibfield  {author} {\bibinfo {author} {\bibfnamefont {B.}~\bibnamefont {Aubert}}, \bibinfo {author} {\bibfnamefont {D.}~\bibnamefont {Boutigny}}, \bibinfo {author} {\bibfnamefont {J.~M.}\ \bibnamefont {Gaillard}}, \bibinfo {author} {\bibfnamefont {A.}~\bibnamefont {Hicheur}}, \bibinfo {author} {\bibfnamefont {Y.}~\bibnamefont {Karyotakis}}, \bibinfo {author} {\bibfnamefont {J.~P.}\ \bibnamefont {Lees}}, \bibinfo {author} {\bibfnamefont {P.}~\bibnamefont {Robbe}}, \bibinfo {author} {\bibfnamefont {V.}~\bibnamefont {Tisserand}}, \bibinfo {author} {\bibfnamefont {A.}~\bibnamefont {Palano}}, \bibinfo {author} {\bibfnamefont {G.~P.}\ \bibnamefont {Chen}},  \emph {et~al.} (\bibinfo {collaboration} {BABAR Collaboration}),\ }\href {\doibase 10.1103/PhysRevLett.87.091801} {\bibfield  {journal} {\bibinfo  {journal} {Phys. Rev. Lett.}\ }\textbf {\bibinfo {volume} {87}},\ \bibinfo {pages} {091801} (\bibinfo {year} {2001})}\BibitemShut {NoStop}%
\bibitem [{\citenamefont {Abe}\ \emph {et~al.}(2001)\citenamefont {Abe}, \citenamefont {Abe}, \citenamefont {Adachi}, \citenamefont {Ahn}, \citenamefont {Aihara}, \citenamefont {Akatsu}, \citenamefont {Alimonti}, \citenamefont {Asai}, \citenamefont {Asai}, \citenamefont {Asano} \emph {et~al.}}]{abe2001observation}%
  \BibitemOpen
  \bibfield  {author} {\bibinfo {author} {\bibfnamefont {K.}~\bibnamefont {Abe}}, \bibinfo {author} {\bibfnamefont {R.}~\bibnamefont {Abe}}, \bibinfo {author} {\bibfnamefont {I.}~\bibnamefont {Adachi}}, \bibinfo {author} {\bibfnamefont {B.~S.}\ \bibnamefont {Ahn}}, \bibinfo {author} {\bibfnamefont {H.}~\bibnamefont {Aihara}}, \bibinfo {author} {\bibfnamefont {M.}~\bibnamefont {Akatsu}}, \bibinfo {author} {\bibfnamefont {G.}~\bibnamefont {Alimonti}}, \bibinfo {author} {\bibfnamefont {K.}~\bibnamefont {Asai}}, \bibinfo {author} {\bibfnamefont {M.}~\bibnamefont {Asai}}, \bibinfo {author} {\bibfnamefont {Y.}~\bibnamefont {Asano}},  \emph {et~al.} (\bibinfo {collaboration} {Belle Collaboration}),\ }\href {\doibase 10.1103/PhysRevLett.87.091802} {\bibfield  {journal} {\bibinfo  {journal} {Phys. Rev. Lett.}\ }\textbf {\bibinfo {volume} {87}},\ \bibinfo {pages} {091802} (\bibinfo {year} {2001})}\BibitemShut {NoStop}%
\bibitem [{\citenamefont {Aaij}\ \emph {et~al.}(2013)\citenamefont {Aaij}, \citenamefont {Beteta}, \citenamefont {Adeva}, \citenamefont {Adinolfi}, \citenamefont {Adrover}, \citenamefont {Affolder}, \citenamefont {Ajaltouni}, \citenamefont {Albrecht}, \citenamefont {Alessio}, \citenamefont {Alexander} \emph {et~al.}}]{aaij2013first}%
  \BibitemOpen
  \bibfield  {author} {\bibinfo {author} {\bibfnamefont {R.}~\bibnamefont {Aaij}}, \bibinfo {author} {\bibfnamefont {C.~A.}\ \bibnamefont {Beteta}}, \bibinfo {author} {\bibfnamefont {B.}~\bibnamefont {Adeva}}, \bibinfo {author} {\bibfnamefont {M.}~\bibnamefont {Adinolfi}}, \bibinfo {author} {\bibfnamefont {C.}~\bibnamefont {Adrover}}, \bibinfo {author} {\bibfnamefont {A.}~\bibnamefont {Affolder}}, \bibinfo {author} {\bibfnamefont {Z.}~\bibnamefont {Ajaltouni}}, \bibinfo {author} {\bibfnamefont {J.}~\bibnamefont {Albrecht}}, \bibinfo {author} {\bibfnamefont {F.}~\bibnamefont {Alessio}}, \bibinfo {author} {\bibfnamefont {M.}~\bibnamefont {Alexander}},  \emph {et~al.} (\bibinfo {collaboration} {LHCb Collaboration}),\ }\href {\doibase 10.1103/PhysRevLett.110.221601} {\bibfield  {journal} {\bibinfo  {journal} {Phys. Rev. Lett.}\ }\textbf {\bibinfo {volume} {110}},\ \bibinfo {pages} {221601} (\bibinfo {year} {2013})}\BibitemShut {NoStop}%
\bibitem [{\citenamefont {Aaij}\ \emph {et~al.}(2019)\citenamefont {Aaij}, \citenamefont {Abell\'an~Beteta}, \citenamefont {Adeva}, \citenamefont {Adinolfi} \emph {et~al.}}]{aaij2019observation}%
  \BibitemOpen
  \bibfield  {author} {\bibinfo {author} {\bibfnamefont {R.}~\bibnamefont {Aaij}}, \bibinfo {author} {\bibfnamefont {C.}~\bibnamefont {Abell\'an~Beteta}}, \bibinfo {author} {\bibfnamefont {B.}~\bibnamefont {Adeva}}, \bibinfo {author} {\bibfnamefont {M.}~\bibnamefont {Adinolfi}},  \emph {et~al.} (\bibinfo {collaboration} {LHCb Collaboration}),\ }\href {\doibase 10.1103/PhysRevLett.122.211803} {\bibfield  {journal} {\bibinfo  {journal} {Phys. Rev. Lett.}\ }\textbf {\bibinfo {volume} {122}},\ \bibinfo {pages} {211803} (\bibinfo {year} {2019})}\BibitemShut {NoStop}%
\bibitem [{\citenamefont {Abe}\ \emph {et~al.}(2020)\citenamefont {Abe}, \citenamefont {Akutsu}, \citenamefont {Ali}, \citenamefont {Alt}, \citenamefont {Andreopoulos}, \citenamefont {Anthony}, \citenamefont {Antonova}, \citenamefont {Aoki}, \citenamefont {Ariga}, \citenamefont {Arihara} \emph {et~al.}}]{t2k2020constraint}%
  \BibitemOpen
  \bibfield  {author} {\bibinfo {author} {\bibfnamefont {K.}~\bibnamefont {Abe}}, \bibinfo {author} {\bibfnamefont {R.}~\bibnamefont {Akutsu}}, \bibinfo {author} {\bibfnamefont {A.}~\bibnamefont {Ali}}, \bibinfo {author} {\bibfnamefont {C.}~\bibnamefont {Alt}}, \bibinfo {author} {\bibfnamefont {C.}~\bibnamefont {Andreopoulos}}, \bibinfo {author} {\bibfnamefont {L.}~\bibnamefont {Anthony}}, \bibinfo {author} {\bibfnamefont {M.}~\bibnamefont {Antonova}}, \bibinfo {author} {\bibfnamefont {S.}~\bibnamefont {Aoki}}, \bibinfo {author} {\bibfnamefont {A.}~\bibnamefont {Ariga}}, \bibinfo {author} {\bibfnamefont {T.}~\bibnamefont {Arihara}},  \emph {et~al.} (\bibinfo {collaboration} {T2K Collaboration}),\ }\href {\doibase 10.1038/s41586-020-2177-0} {\bibfield  {journal} {\bibinfo  {journal} {Nature}\ }\textbf {\bibinfo {volume} {580}},\ \bibinfo {pages} {339} (\bibinfo {year} {2020})}\BibitemShut {NoStop}%
\bibitem [{t2k(2020)}]{t2k2020publisher}%
  \BibitemOpen
  \href {\doibase 10.1038/s41586-020-2415-5} {\bibfield  {journal} {\bibinfo  {journal} {Nature}\ }\textbf {\bibinfo {volume} {583}},\ \bibinfo {pages} {E16} (\bibinfo {year} {2020})}\BibitemShut {NoStop}%
\bibitem [{\citenamefont {Abe}\ \emph {et~al.}(2023)\citenamefont {Abe}, \citenamefont {Akhlaq}, \citenamefont {Akutsu}, \citenamefont {Ali}, \citenamefont {Monsalve}, \citenamefont {Alt}, \citenamefont {Andreopoulos}, \citenamefont {Antonova}, \citenamefont {Aoki}, \citenamefont {Arihara} \emph {et~al.}}]{abe2023measurements}%
  \BibitemOpen
  \bibfield  {author} {\bibinfo {author} {\bibfnamefont {K.}~\bibnamefont {Abe}}, \bibinfo {author} {\bibfnamefont {N.}~\bibnamefont {Akhlaq}}, \bibinfo {author} {\bibfnamefont {R.}~\bibnamefont {Akutsu}}, \bibinfo {author} {\bibfnamefont {A.}~\bibnamefont {Ali}}, \bibinfo {author} {\bibfnamefont {S.~A.}\ \bibnamefont {Monsalve}}, \bibinfo {author} {\bibfnamefont {C.}~\bibnamefont {Alt}}, \bibinfo {author} {\bibfnamefont {C.}~\bibnamefont {Andreopoulos}}, \bibinfo {author} {\bibfnamefont {M.}~\bibnamefont {Antonova}}, \bibinfo {author} {\bibfnamefont {S.}~\bibnamefont {Aoki}}, \bibinfo {author} {\bibfnamefont {T.}~\bibnamefont {Arihara}},  \emph {et~al.} (\bibinfo {collaboration} {T2K Collaboration}),\ }\href {\doibase 10.1140/epjc/s10052-023-11819-x} {\bibfield  {journal} {\bibinfo  {journal} {Eur. Phys. J. C}\ }\textbf {\bibinfo {volume} {83}},\ \bibinfo {pages} {782} (\bibinfo {year} {2023})}\BibitemShut {NoStop}%
\bibitem [{\citenamefont {Sakharov}(1967{\natexlab{a}})}]{sakharov1967violation}%
  \BibitemOpen
  \bibfield  {author} {\bibinfo {author} {\bibfnamefont {A.~D.}\ \bibnamefont {Sakharov}},\ }\href@noop {} {\bibfield  {journal} {\bibinfo  {journal} {JETP Lett.}\ }\textbf {\bibinfo {volume} {5}},\ \bibinfo {pages} {24} (\bibinfo {year} {1967}{\natexlab{a}})}\BibitemShut {NoStop}%
\bibitem [{\citenamefont {Sakharov}(1967{\natexlab{b}})}]{sakharov1967violation2}%
  \BibitemOpen
  \bibfield  {author} {\bibinfo {author} {\bibfnamefont {A.~D.}\ \bibnamefont {Sakharov}},\ }\href {\doibase 10.1070/PU1991v034n05ABEH002497} {\bibfield  {journal} {\bibinfo  {journal} {Sov. Phys. Usp.}\ }\textbf {\bibinfo {volume} {34}},\ \bibinfo {pages} {392} (\bibinfo {year} {1967}{\natexlab{b}})}\BibitemShut {NoStop}%
\bibitem [{\citenamefont {Bjorken}\ \emph {et~al.}(1964)\citenamefont {Bjorken}, \citenamefont {Drell},\ and\ \citenamefont {Mansfield}}]{bjorken1964relativistic}%
  \BibitemOpen
  \bibfield  {author} {\bibinfo {author} {\bibfnamefont {J.~D.}\ \bibnamefont {Bjorken}}, \bibinfo {author} {\bibfnamefont {S.~D.}\ \bibnamefont {Drell}}, \ and\ \bibinfo {author} {\bibfnamefont {J.~E.}\ \bibnamefont {Mansfield}},\ }\href@noop {} {\emph {\bibinfo {title} {Relativistic quantum mechanics}}}\ (\bibinfo  {publisher} {McGraw-Hill},\ \bibinfo {address} {New York},\ \bibinfo {year} {1964})\BibitemShut {NoStop}%
\bibitem [{\citenamefont {McKeen}\ and\ \citenamefont {Nelson}(2016)}]{mckeen2016c}%
  \BibitemOpen
  \bibfield  {author} {\bibinfo {author} {\bibfnamefont {D.}~\bibnamefont {McKeen}}\ and\ \bibinfo {author} {\bibfnamefont {A.~E.}\ \bibnamefont {Nelson}},\ }\href {\doibase 10.1103/PhysRevD.94.076002} {\bibfield  {journal} {\bibinfo  {journal} {Phys. Rev. D}\ }\textbf {\bibinfo {volume} {94}},\ \bibinfo {pages} {076002} (\bibinfo {year} {2016})}\BibitemShut {NoStop}%
\bibitem [{\citenamefont {Nussinov}\ and\ \citenamefont {Shrock}(2002)}]{nussinov2002n}%
  \BibitemOpen
  \bibfield  {author} {\bibinfo {author} {\bibfnamefont {S.}~\bibnamefont {Nussinov}}\ and\ \bibinfo {author} {\bibfnamefont {R.}~\bibnamefont {Shrock}},\ }\href {\doibase 10.1103/PhysRevLett.88.171601} {\bibfield  {journal} {\bibinfo  {journal} {Phys. Rev. Lett.}\ }\textbf {\bibinfo {volume} {88}},\ \bibinfo {pages} {171601} (\bibinfo {year} {2002})}\BibitemShut {NoStop}%
\bibitem [{\citenamefont {Girmohanta}\ and\ \citenamefont {Shrock}(2020)}]{girmohanta2020nucleon}%
  \BibitemOpen
  \bibfield  {author} {\bibinfo {author} {\bibfnamefont {S.}~\bibnamefont {Girmohanta}}\ and\ \bibinfo {author} {\bibfnamefont {R.}~\bibnamefont {Shrock}},\ }\href {\doibase 10.1103/PhysRevD.101.095012} {\bibfield  {journal} {\bibinfo  {journal} {Phys. Rev. D}\ }\textbf {\bibinfo {volume} {101}},\ \bibinfo {pages} {095012} (\bibinfo {year} {2020})}\BibitemShut {NoStop}%
\bibitem [{\citenamefont {Fornal}\ and\ \citenamefont {Grinstein}(2018)}]{fornal2018dark}%
  \BibitemOpen
  \bibfield  {author} {\bibinfo {author} {\bibfnamefont {B.}~\bibnamefont {Fornal}}\ and\ \bibinfo {author} {\bibfnamefont {B.}~\bibnamefont {Grinstein}},\ }\href {\doibase 10.1103/PhysRevLett.120.191801} {\bibfield  {journal} {\bibinfo  {journal} {Phys. Rev. Lett.}\ }\textbf {\bibinfo {volume} {120}},\ \bibinfo {pages} {191801} (\bibinfo {year} {2018})}\BibitemShut {NoStop}%
\bibitem [{\citenamefont {Fornal}(2023)}]{fornal2023neutron}%
  \BibitemOpen
  \bibfield  {author} {\bibinfo {author} {\bibfnamefont {B.}~\bibnamefont {Fornal}},\ }\href {\doibase 10.3390/universe9100449} {\bibfield  {journal} {\bibinfo  {journal} {Universe}\ }\textbf {\bibinfo {volume} {9}},\ \bibinfo {pages} {449} (\bibinfo {year} {2023})}\BibitemShut {NoStop}%
\bibitem [{\citenamefont {McKeen}\ \emph {et~al.}(2018)\citenamefont {McKeen}, \citenamefont {Nelson}, \citenamefont {Reddy},\ and\ \citenamefont {Zhou}}]{mckeen2018neutron}%
  \BibitemOpen
  \bibfield  {author} {\bibinfo {author} {\bibfnamefont {D.}~\bibnamefont {McKeen}}, \bibinfo {author} {\bibfnamefont {A.~E.}\ \bibnamefont {Nelson}}, \bibinfo {author} {\bibfnamefont {S.}~\bibnamefont {Reddy}}, \ and\ \bibinfo {author} {\bibfnamefont {D.}~\bibnamefont {Zhou}},\ }\href {\doibase 10.1103/PhysRevLett.121.061802} {\bibfield  {journal} {\bibinfo  {journal} {Phys. Rev. Lett.}\ }\textbf {\bibinfo {volume} {121}},\ \bibinfo {pages} {061802} (\bibinfo {year} {2018})}\BibitemShut {NoStop}%
\bibitem [{\citenamefont {Tiesinga}\ \emph {et~al.}(2021{\natexlab{a}})\citenamefont {Tiesinga}, \citenamefont {Mohr}, \citenamefont {Newell},\ and\ \citenamefont {Taylor}}]{tiesinga2021codata}%
  \BibitemOpen
  \bibfield  {author} {\bibinfo {author} {\bibfnamefont {E.}~\bibnamefont {Tiesinga}}, \bibinfo {author} {\bibfnamefont {P.~J.}\ \bibnamefont {Mohr}}, \bibinfo {author} {\bibfnamefont {D.~B.}\ \bibnamefont {Newell}}, \ and\ \bibinfo {author} {\bibfnamefont {B.~N.}\ \bibnamefont {Taylor}},\ }\href {\doibase 10.1063/5.0064853} {\bibfield  {journal} {\bibinfo  {journal} {J. Phys. Chem. Ref. Dat.}\ }\textbf {\bibinfo {volume} {50}},\ \bibinfo {pages} {033105} (\bibinfo {year} {2021}{\natexlab{a}})}\BibitemShut {NoStop}%
\bibitem [{\citenamefont {Tiesinga}\ \emph {et~al.}(2021{\natexlab{b}})\citenamefont {Tiesinga}, \citenamefont {Mohr}, \citenamefont {Newell},\ and\ \citenamefont {Taylor}}]{tiesinga2021codata2}%
  \BibitemOpen
  \bibfield  {author} {\bibinfo {author} {\bibfnamefont {E.}~\bibnamefont {Tiesinga}}, \bibinfo {author} {\bibfnamefont {P.~J.}\ \bibnamefont {Mohr}}, \bibinfo {author} {\bibfnamefont {D.~B.}\ \bibnamefont {Newell}}, \ and\ \bibinfo {author} {\bibfnamefont {B.~N.}\ \bibnamefont {Taylor}},\ }\href {\doibase 10.1103/RevModPhys.93.025010} {\bibfield  {journal} {\bibinfo  {journal} {Rev. Mod. Phys.}\ }\textbf {\bibinfo {volume} {93}},\ \bibinfo {pages} {025010} (\bibinfo {year} {2021}{\natexlab{b}})}\BibitemShut {NoStop}%
\bibitem [{\citenamefont {Allahverdi}\ and\ \citenamefont {Dutta}(2013)}]{allahverdi2013natural}%
  \BibitemOpen
  \bibfield  {author} {\bibinfo {author} {\bibfnamefont {R.}~\bibnamefont {Allahverdi}}\ and\ \bibinfo {author} {\bibfnamefont {B.}~\bibnamefont {Dutta}},\ }\href {\doibase 10.1103/PhysRevD.88.023525} {\bibfield  {journal} {\bibinfo  {journal} {Phys. Rev. D}\ }\textbf {\bibinfo {volume} {88}},\ \bibinfo {pages} {023525} (\bibinfo {year} {2013})}\BibitemShut {NoStop}%
\bibitem [{\citenamefont {Allahverdi}\ \emph {et~al.}(2014)\citenamefont {Allahverdi}, \citenamefont {Dutta},\ and\ \citenamefont {Gao}}]{allahverdi2014kev}%
  \BibitemOpen
  \bibfield  {author} {\bibinfo {author} {\bibfnamefont {R.}~\bibnamefont {Allahverdi}}, \bibinfo {author} {\bibfnamefont {B.}~\bibnamefont {Dutta}}, \ and\ \bibinfo {author} {\bibfnamefont {Y.}~\bibnamefont {Gao}},\ }\href {\doibase 10.1103/PhysRevD.89.127305} {\bibfield  {journal} {\bibinfo  {journal} {Phys. Rev. D}\ }\textbf {\bibinfo {volume} {89}},\ \bibinfo {pages} {127305} (\bibinfo {year} {2014})}\BibitemShut {NoStop}%
\bibitem [{\citenamefont {Allahverdi}\ \emph {et~al.}(2018)\citenamefont {Allahverdi}, \citenamefont {Dev},\ and\ \citenamefont {Dutta}}]{allahverdi2018simple}%
  \BibitemOpen
  \bibfield  {author} {\bibinfo {author} {\bibfnamefont {R.}~\bibnamefont {Allahverdi}}, \bibinfo {author} {\bibfnamefont {P.~S.~B.}\ \bibnamefont {Dev}}, \ and\ \bibinfo {author} {\bibfnamefont {B.}~\bibnamefont {Dutta}},\ }\href {\doibase 10.1016/j.physletb.2018.02.019} {\bibfield  {journal} {\bibinfo  {journal} {Phys. Lett. B}\ }\textbf {\bibinfo {volume} {779}},\ \bibinfo {pages} {262} (\bibinfo {year} {2018})}\BibitemShut {NoStop}%
\bibitem [{\citenamefont {Jin}\ and\ \citenamefont {Gao}(2018)}]{jin2018nucleon}%
  \BibitemOpen
  \bibfield  {author} {\bibinfo {author} {\bibfnamefont {M.}~\bibnamefont {Jin}}\ and\ \bibinfo {author} {\bibfnamefont {Y.}~\bibnamefont {Gao}},\ }\href {\doibase 10.1103/PhysRevD.98.075026} {\bibfield  {journal} {\bibinfo  {journal} {Phys. Rev. D}\ }\textbf {\bibinfo {volume} {98}},\ \bibinfo {pages} {075026} (\bibinfo {year} {2018})}\BibitemShut {NoStop}%
\bibitem [{\citenamefont {Fajfer}\ and\ \citenamefont {Susi{\v{c}}}(2021)}]{fajfer2021colored}%
  \BibitemOpen
  \bibfield  {author} {\bibinfo {author} {\bibfnamefont {S.}~\bibnamefont {Fajfer}}\ and\ \bibinfo {author} {\bibfnamefont {D.}~\bibnamefont {Susi{\v{c}}}},\ }\href {\doibase 10.1103/PhysRevD.103.055012} {\bibfield  {journal} {\bibinfo  {journal} {Phys. Rev. D}\ }\textbf {\bibinfo {volume} {103}},\ \bibinfo {pages} {055012} (\bibinfo {year} {2021})}\BibitemShut {NoStop}%
\bibitem [{\citenamefont {McKeen}\ and\ \citenamefont {Pospelov}(2023)}]{mckeen2023long}%
  \BibitemOpen
  \bibfield  {author} {\bibinfo {author} {\bibfnamefont {D.}~\bibnamefont {McKeen}}\ and\ \bibinfo {author} {\bibfnamefont {M.}~\bibnamefont {Pospelov}},\ }\href {\doibase 10.3390/universe9110473} {\bibfield  {journal} {\bibinfo  {journal} {Universe}\ }\textbf {\bibinfo {volume} {9}},\ \bibinfo {pages} {473} (\bibinfo {year} {2023})}\BibitemShut {NoStop}%
\bibitem [{\citenamefont {Gu}\ and\ \citenamefont {Sarkar}(2011)}]{gu2011baryogenesis}%
  \BibitemOpen
  \bibfield  {author} {\bibinfo {author} {\bibfnamefont {P.}~\bibnamefont {Gu}}\ and\ \bibinfo {author} {\bibfnamefont {U.}~\bibnamefont {Sarkar}},\ }\href {\doibase 10.1016/j.physletb.2011.10.017} {\bibfield  {journal} {\bibinfo  {journal} {Phys. Lett. B}\ }\textbf {\bibinfo {volume} {705}},\ \bibinfo {pages} {170} (\bibinfo {year} {2011})}\BibitemShut {NoStop}%
\bibitem [{\citenamefont {Dev}\ and\ \citenamefont {Mohapatra}(2015)}]{dev2015tev}%
  \BibitemOpen
  \bibfield  {author} {\bibinfo {author} {\bibfnamefont {P.~S.~B.}\ \bibnamefont {Dev}}\ and\ \bibinfo {author} {\bibfnamefont {R.~N.}\ \bibnamefont {Mohapatra}},\ }\href {\doibase 10.1103/PhysRevD.92.016007} {\bibfield  {journal} {\bibinfo  {journal} {Phys. Rev. D}\ }\textbf {\bibinfo {volume} {92}},\ \bibinfo {pages} {016007} (\bibinfo {year} {2015})}\BibitemShut {NoStop}%
\bibitem [{\citenamefont {Aoki}\ \emph {et~al.}(2017)\citenamefont {Aoki}, \citenamefont {Izubuchi}, \citenamefont {Shintani},\ and\ \citenamefont {Soni}}]{aoki2017improved}%
  \BibitemOpen
  \bibfield  {author} {\bibinfo {author} {\bibfnamefont {Y.}~\bibnamefont {Aoki}}, \bibinfo {author} {\bibfnamefont {T.}~\bibnamefont {Izubuchi}}, \bibinfo {author} {\bibfnamefont {E.}~\bibnamefont {Shintani}}, \ and\ \bibinfo {author} {\bibfnamefont {A.}~\bibnamefont {Soni}},\ }\href {\doibase 10.1103/PhysRevD.96.014506} {\bibfield  {journal} {\bibinfo  {journal} {Phys. Rev. D}\ }\textbf {\bibinfo {volume} {96}},\ \bibinfo {pages} {014506} (\bibinfo {year} {2017})}\BibitemShut {NoStop}%
\bibitem [{\citenamefont {Mohapatra}\ \emph {et~al.}(2008)\citenamefont {Mohapatra}, \citenamefont {Okada},\ and\ \citenamefont {Yu}}]{mohapatra2008diquark}%
  \BibitemOpen
  \bibfield  {author} {\bibinfo {author} {\bibfnamefont {R.~N.}\ \bibnamefont {Mohapatra}}, \bibinfo {author} {\bibfnamefont {N.}~\bibnamefont {Okada}}, \ and\ \bibinfo {author} {\bibfnamefont {H.~B.}\ \bibnamefont {Yu}},\ }\href {\doibase 10.1103/PhysRevD.77.011701} {\bibfield  {journal} {\bibinfo  {journal} {Phys. Rev. D}\ }\textbf {\bibinfo {volume} {77}},\ \bibinfo {pages} {011701} (\bibinfo {year} {2008})}\BibitemShut {NoStop}%
\bibitem [{\citenamefont {Babu}\ \emph {et~al.}(2009)\citenamefont {Babu}, \citenamefont {Dev},\ and\ \citenamefont {Mohapatra}}]{babu2009neutrino}%
  \BibitemOpen
  \bibfield  {author} {\bibinfo {author} {\bibfnamefont {K.~S.}\ \bibnamefont {Babu}}, \bibinfo {author} {\bibfnamefont {P.~S.~B.}\ \bibnamefont {Dev}}, \ and\ \bibinfo {author} {\bibfnamefont {R.~N.}\ \bibnamefont {Mohapatra}},\ }\href {\doibase 10.1103/PhysRevD.79.015017} {\bibfield  {journal} {\bibinfo  {journal} {Phys. Rev. D}\ }\textbf {\bibinfo {volume} {79}},\ \bibinfo {pages} {015017} (\bibinfo {year} {2009})}\BibitemShut {NoStop}%
\bibitem [{\citenamefont {Saha}\ \emph {et~al.}(2010)\citenamefont {Saha}, \citenamefont {Misra},\ and\ \citenamefont {Kundu}}]{saha2010constraining}%
  \BibitemOpen
  \bibfield  {author} {\bibinfo {author} {\bibfnamefont {J.~P.}\ \bibnamefont {Saha}}, \bibinfo {author} {\bibfnamefont {B.}~\bibnamefont {Misra}}, \ and\ \bibinfo {author} {\bibfnamefont {A.}~\bibnamefont {Kundu}},\ }\href {\doibase 10.1103/PhysRevD.81.095011} {\bibfield  {journal} {\bibinfo  {journal} {Phys. Rev. D}\ }\textbf {\bibinfo {volume} {81}},\ \bibinfo {pages} {095011} (\bibinfo {year} {2010})}\BibitemShut {NoStop}%
\bibitem [{\citenamefont {Dor{\v{s}}ner}\ \emph {et~al.}(2010)\citenamefont {Dor{\v{s}}ner}, \citenamefont {Fajfer}, \citenamefont {Kamenik},\ and\ \citenamefont {Ko{\v{s}}nik}}]{dorvsner2010light}%
  \BibitemOpen
  \bibfield  {author} {\bibinfo {author} {\bibfnamefont {I.}~\bibnamefont {Dor{\v{s}}ner}}, \bibinfo {author} {\bibfnamefont {S.}~\bibnamefont {Fajfer}}, \bibinfo {author} {\bibfnamefont {J.~F.}\ \bibnamefont {Kamenik}}, \ and\ \bibinfo {author} {\bibfnamefont {N.}~\bibnamefont {Ko{\v{s}}nik}},\ }\href {\doibase 10.1103/PhysRevD.82.094015} {\bibfield  {journal} {\bibinfo  {journal} {Phys. Rev. D}\ }\textbf {\bibinfo {volume} {82}},\ \bibinfo {pages} {094015} (\bibinfo {year} {2010})}\BibitemShut {NoStop}%
\bibitem [{\citenamefont {Giudice}\ \emph {et~al.}(2011)\citenamefont {Giudice}, \citenamefont {G.},\ and\ \citenamefont {Sundrum}}]{giudice2011flavourful}%
  \BibitemOpen
  \bibfield  {author} {\bibinfo {author} {\bibfnamefont {G.~F.}\ \bibnamefont {Giudice}}, \bibinfo {author} {\bibfnamefont {B.}~\bibnamefont {G.}}, \ and\ \bibinfo {author} {\bibfnamefont {R.}~\bibnamefont {Sundrum}},\ }\href {\doibase 10.1007/JHEP08(2011)055} {\bibfield  {journal} {\bibinfo  {journal} {J. High Energy Phys.}\ }\textbf {\bibinfo {volume} {08}},\ \bibinfo {pages} {055} (\bibinfo {year} {2011})}\BibitemShut {NoStop}%
\bibitem [{\citenamefont {Dor{\v{s}}ner}\ \emph {et~al.}(2011)\citenamefont {Dor{\v{s}}ner}, \citenamefont {Drobnak}, \citenamefont {Fajfer}, \citenamefont {Kamenik},\ and\ \citenamefont {Ko{\v{s}}nik}}]{dorvsner2011limits}%
  \BibitemOpen
  \bibfield  {author} {\bibinfo {author} {\bibfnamefont {I.}~\bibnamefont {Dor{\v{s}}ner}}, \bibinfo {author} {\bibfnamefont {J.}~\bibnamefont {Drobnak}}, \bibinfo {author} {\bibfnamefont {S.}~\bibnamefont {Fajfer}}, \bibinfo {author} {\bibfnamefont {J.~F.}\ \bibnamefont {Kamenik}}, \ and\ \bibinfo {author} {\bibfnamefont {N.}~\bibnamefont {Ko{\v{s}}nik}},\ }\href {\doibase 10.1007/JHEP11(2011)002} {\bibfield  {journal} {\bibinfo  {journal} {J. High Energ. Phys.}\ }\textbf {\bibinfo {volume} {11}},\ \bibinfo {pages} {002} (\bibinfo {year} {2011})}\BibitemShut {NoStop}%
\bibitem [{\citenamefont {Barr}\ and\ \citenamefont {Calmet}(2012)}]{barr2012observable}%
  \BibitemOpen
  \bibfield  {author} {\bibinfo {author} {\bibfnamefont {S.~M.}\ \bibnamefont {Barr}}\ and\ \bibinfo {author} {\bibfnamefont {X.}~\bibnamefont {Calmet}},\ }\href {\doibase 10.1103/PhysRevD.86.116010} {\bibfield  {journal} {\bibinfo  {journal} {Phys. Rev. D}\ }\textbf {\bibinfo {volume} {86}},\ \bibinfo {pages} {116010} (\bibinfo {year} {2012})}\BibitemShut {NoStop}%
\bibitem [{\citenamefont {Babu}\ \emph {et~al.}(2013{\natexlab{a}})\citenamefont {Babu}, \citenamefont {Dev}, \citenamefont {Fortes},\ and\ \citenamefont {Mohapatra}}]{babu2013post}%
  \BibitemOpen
  \bibfield  {author} {\bibinfo {author} {\bibfnamefont {K.~S.}\ \bibnamefont {Babu}}, \bibinfo {author} {\bibfnamefont {P.~S.~B.}\ \bibnamefont {Dev}}, \bibinfo {author} {\bibfnamefont {E.~C.}\ \bibnamefont {Fortes}}, \ and\ \bibinfo {author} {\bibfnamefont {R.~N.}\ \bibnamefont {Mohapatra}},\ }\href {\doibase 10.1103/PhysRevD.87.115019} {\bibfield  {journal} {\bibinfo  {journal} {Phys. Rev. D}\ }\textbf {\bibinfo {volume} {87}},\ \bibinfo {pages} {115019} (\bibinfo {year} {2013}{\natexlab{a}})}\BibitemShut {NoStop}%
\bibitem [{\citenamefont {Babu}\ \emph {et~al.}(2013{\natexlab{b}})\citenamefont {Babu}, \citenamefont {Dev}, \citenamefont {Fortes},\ and\ \citenamefont {Mohapatra}}]{babu2013expectations}%
  \BibitemOpen
  \bibfield  {author} {\bibinfo {author} {\bibfnamefont {K.~S.}\ \bibnamefont {Babu}}, \bibinfo {author} {\bibfnamefont {P.~S.~B.}\ \bibnamefont {Dev}}, \bibinfo {author} {\bibfnamefont {E.~C.}\ \bibnamefont {Fortes}}, \ and\ \bibinfo {author} {\bibfnamefont {R.~N.}\ \bibnamefont {Mohapatra}},\ }\href {\doibase 10.1063/1.4807359} {\bibfield  {journal} {\bibinfo  {journal} {AIP Conf. Proc.}\ }\textbf {\bibinfo {volume} {1534}},\ \bibinfo {pages} {211} (\bibinfo {year} {2013}{\natexlab{b}})}\BibitemShut {NoStop}%
\bibitem [{\citenamefont {Fortes}\ \emph {et~al.}(2013)\citenamefont {Fortes}, \citenamefont {Babu},\ and\ \citenamefont {Mohapatra}}]{fortes2013flavor}%
  \BibitemOpen
  \bibfield  {author} {\bibinfo {author} {\bibfnamefont {E.~C.}\ \bibnamefont {Fortes}}, \bibinfo {author} {\bibfnamefont {K.~S.}\ \bibnamefont {Babu}}, \ and\ \bibinfo {author} {\bibfnamefont {R.~N.}\ \bibnamefont {Mohapatra}},\ }\href {https://arxiv.org/abs/1311.4101} {\bibfield  {journal} {\bibinfo  {journal} {arXiv:1311.4101}\ } (\bibinfo {year} {2013})}\BibitemShut {NoStop}%
\bibitem [{\citenamefont {Patra}\ and\ \citenamefont {Pritimita}(2014)}]{patra2014post}%
  \BibitemOpen
  \bibfield  {author} {\bibinfo {author} {\bibfnamefont {S.}~\bibnamefont {Patra}}\ and\ \bibinfo {author} {\bibfnamefont {P.}~\bibnamefont {Pritimita}},\ }\href {\doibase 10.1140/epjc/s10052-014-3078-x} {\bibfield  {journal} {\bibinfo  {journal} {Eur. Phys. J. C}\ }\textbf {\bibinfo {volume} {74}},\ \bibinfo {pages} {3078} (\bibinfo {year} {2014})}\BibitemShut {NoStop}%
\bibitem [{\citenamefont {Sahoo}\ and\ \citenamefont {Mohanta}(2015)}]{sahoo2015scalar}%
  \BibitemOpen
  \bibfield  {author} {\bibinfo {author} {\bibfnamefont {S.}~\bibnamefont {Sahoo}}\ and\ \bibinfo {author} {\bibfnamefont {R.}~\bibnamefont {Mohanta}},\ }\href {\doibase 10.1103/PhysRevD.91.094019} {\bibfield  {journal} {\bibinfo  {journal} {Phys. Rev. D}\ }\textbf {\bibinfo {volume} {91}},\ \bibinfo {pages} {094019} (\bibinfo {year} {2015})}\BibitemShut {NoStop}%
\bibitem [{\citenamefont {Addazi}(2015)}]{addazi2015exotic}%
  \BibitemOpen
  \bibfield  {author} {\bibinfo {author} {\bibfnamefont {A.}~\bibnamefont {Addazi}},\ }\href {\doibase 10.1007/JHEP04(2015)153} {\bibfield  {journal} {\bibinfo  {journal} {J. High Energ. Phys.}\ }\textbf {\bibinfo {volume} {04}},\ \bibinfo {pages} {153} (\bibinfo {year} {2015})}\BibitemShut {NoStop}%
\bibitem [{\citenamefont {Kim}\ \emph {et~al.}(2019)\citenamefont {Kim}, \citenamefont {Ko}, \citenamefont {Li}, \citenamefont {Park},\ and\ \citenamefont {Wu}}]{kim2019correlation}%
  \BibitemOpen
  \bibfield  {author} {\bibinfo {author} {\bibfnamefont {T.~J.}\ \bibnamefont {Kim}}, \bibinfo {author} {\bibfnamefont {P.}~\bibnamefont {Ko}}, \bibinfo {author} {\bibfnamefont {J.}~\bibnamefont {Li}}, \bibinfo {author} {\bibfnamefont {J.}~\bibnamefont {Park}}, \ and\ \bibinfo {author} {\bibfnamefont {P.}~\bibnamefont {Wu}},\ }\href {\doibase 10.1007/JHEP07(2019)025} {\bibfield  {journal} {\bibinfo  {journal} {J. High Energ. Phys.}\ }\textbf {\bibinfo {volume} {07}},\ \bibinfo {pages} {025} (\bibinfo {year} {2019})}\BibitemShut {NoStop}%
\bibitem [{\citenamefont {Fridell}\ \emph {et~al.}(2021)\citenamefont {Fridell}, \citenamefont {Harz},\ and\ \citenamefont {Hati}}]{fridell2021probing}%
  \BibitemOpen
  \bibfield  {author} {\bibinfo {author} {\bibfnamefont {K.}~\bibnamefont {Fridell}}, \bibinfo {author} {\bibfnamefont {J.}~\bibnamefont {Harz}}, \ and\ \bibinfo {author} {\bibfnamefont {C.}~\bibnamefont {Hati}},\ }\href {\doibase 10.1007/JHEP11(2021)185} {\bibfield  {journal} {\bibinfo  {journal} {J. High Energ. Phys.}\ }\textbf {\bibinfo {volume} {11}},\ \bibinfo {pages} {185} (\bibinfo {year} {2021})}\BibitemShut {NoStop}%
\bibitem [{\citenamefont {Zhou}\ and\ \citenamefont {Tang}(2023)}]{zhou2023analysis}%
  \BibitemOpen
  \bibfield  {author} {\bibinfo {author} {\bibfnamefont {C.}~\bibnamefont {Zhou}}\ and\ \bibinfo {author} {\bibfnamefont {J.}~\bibnamefont {Tang}},\ }\href {\doibase 10.1016/j.csite.2023.102868} {\bibfield  {journal} {\bibinfo  {journal} {Cas. Stud. Therm. Eng.}\ }\textbf {\bibinfo {volume} {44}},\ \bibinfo {pages} {102868} (\bibinfo {year} {2023})}\BibitemShut {NoStop}%
\bibitem [{\citenamefont {Gudkov}\ \emph {et~al.}(2020)\citenamefont {Gudkov}, \citenamefont {Nesvizhevsky}, \citenamefont {Protasov}, \citenamefont {Snow},\ and\ \citenamefont {Voronin}}]{gudkov2020new}%
  \BibitemOpen
  \bibfield  {author} {\bibinfo {author} {\bibfnamefont {V.}~\bibnamefont {Gudkov}}, \bibinfo {author} {\bibfnamefont {V.~V.}\ \bibnamefont {Nesvizhevsky}}, \bibinfo {author} {\bibfnamefont {K.~V.}\ \bibnamefont {Protasov}}, \bibinfo {author} {\bibfnamefont {W.~M.}\ \bibnamefont {Snow}}, \ and\ \bibinfo {author} {\bibfnamefont {A.~Y.}\ \bibnamefont {Voronin}},\ }\href {\doibase 10.1016/j.physletb.2020.135636} {\bibfield  {journal} {\bibinfo  {journal} {Phys. Lett. B}\ }\textbf {\bibinfo {volume} {808}},\ \bibinfo {pages} {135636} (\bibinfo {year} {2020})}\BibitemShut {NoStop}%
\bibitem [{\citenamefont {Ayres}\ \emph {et~al.}(2022{\natexlab{b}})\citenamefont {Ayres}, \citenamefont {Ban}, \citenamefont {Bison}, \citenamefont {Bodek}, \citenamefont {Bondar}, \citenamefont {Bouillaud}, \citenamefont {Clement}, \citenamefont {Chanel}, \citenamefont {Chiu}, \citenamefont {Crawford} \emph {et~al.}}]{ayres2022very}%
  \BibitemOpen
  \bibfield  {author} {\bibinfo {author} {\bibfnamefont {N.~J.}\ \bibnamefont {Ayres}}, \bibinfo {author} {\bibfnamefont {G.}~\bibnamefont {Ban}}, \bibinfo {author} {\bibfnamefont {G.}~\bibnamefont {Bison}}, \bibinfo {author} {\bibfnamefont {K.}~\bibnamefont {Bodek}}, \bibinfo {author} {\bibfnamefont {V.}~\bibnamefont {Bondar}}, \bibinfo {author} {\bibfnamefont {T.}~\bibnamefont {Bouillaud}}, \bibinfo {author} {\bibfnamefont {B.}~\bibnamefont {Clement}}, \bibinfo {author} {\bibfnamefont {E.}~\bibnamefont {Chanel}}, \bibinfo {author} {\bibfnamefont {P.~J.}\ \bibnamefont {Chiu}}, \bibinfo {author} {\bibfnamefont {C.~B.}\ \bibnamefont {Crawford}},  \emph {et~al.},\ }\href {\doibase 10.1063/5.0101391} {\bibfield  {journal} {\bibinfo  {journal} {Rev. Sci. Instrum.}\ }\textbf {\bibinfo {volume} {93}},\ \bibinfo {pages} {095105} (\bibinfo {year} {2022}{\natexlab{b}})}\BibitemShut {NoStop}%
\bibitem [{\citenamefont {Mereghetti}\ \emph {et~al.}(2015)\citenamefont {Mereghetti}, \citenamefont {Pons},\ and\ \citenamefont {Melatos}}]{mereghetti2015magnetars}%
  \BibitemOpen
  \bibfield  {author} {\bibinfo {author} {\bibfnamefont {S.}~\bibnamefont {Mereghetti}}, \bibinfo {author} {\bibfnamefont {J.~A.}\ \bibnamefont {Pons}}, \ and\ \bibinfo {author} {\bibfnamefont {A.}~\bibnamefont {Melatos}},\ }\href {\doibase 10.1007/s11214-015-0146-y} {\bibfield  {journal} {\bibinfo  {journal} {Space Sci. Rev.}\ }\textbf {\bibinfo {volume} {191}},\ \bibinfo {pages} {315} (\bibinfo {year} {2015})}\BibitemShut {NoStop}%
\bibitem [{\citenamefont {Berezhiani}\ and\ \citenamefont {Vainshtein}(2015)}]{berezhiani2015neutronantineutron}%
  \BibitemOpen
  \bibfield  {author} {\bibinfo {author} {\bibfnamefont {Z.}~\bibnamefont {Berezhiani}}\ and\ \bibinfo {author} {\bibfnamefont {A.}~\bibnamefont {Vainshtein}},\ }\href@noop {} {\enquote {\bibinfo {title} {Neutron-antineutron oscillation as a signal of cp violation},}\ } (\bibinfo {year} {2015})\BibitemShut {NoStop}%
\bibitem [{\citenamefont {Fujikawa}\ and\ \citenamefont {Tureanu}(2015)}]{fujikawa2015neutronantineutron}%
  \BibitemOpen
  \bibfield  {author} {\bibinfo {author} {\bibfnamefont {K.}~\bibnamefont {Fujikawa}}\ and\ \bibinfo {author} {\bibfnamefont {A.}~\bibnamefont {Tureanu}},\ }\href@noop {} {\enquote {\bibinfo {title} {Neutron-antineutron oscillation and parity and cp symmetries},}\ } (\bibinfo {year} {2015})\BibitemShut {NoStop}%
\end{thebibliography}%
\end{document}